# Hologram generation using attention for Fraunhofer diffraction

**Authors:** Dilawer Singh[1*], Antoni J. Wojcik[1], Timothy D. Wilkinson[1]

**Affiliations:**

[1]Electrical Engineering, University of Cambridge, 9 JJ Thompson Ave., Cambridge, CB3 0FA, United Kingdom.

*Correspondence to: ds2070@cam.ac.uk

**Summary Paragraph:** Controlling optical wavefronts in the strict far-field regime is a fully global inverse problem. Fraunhofer propagation is a single Fourier transform of the aperture, so every modulator pixel couples to every point of the reconstruction and no spatial locality survives for the convolutional priors that dominate learned holography to exploit. Iterative phase retrieval stagnates and trades computation for quality, and camera-in-the-loop methods remain iterative and near-field. Here we show that this global coupling maps naturally onto self-attention, generating a phase-only hologram in a single forward pass of 12 to 24 ms. On hardware, every simulation-trained method collapses to a common floor between 15.8 and 16.1 dB median PSNR set by model-reality mismatch, despite spanning 27.7 to 56.5 dB median PSNR in ideal simulation. Placing the physical optics inside the training loop, which we call self-attention in-the-loop (SAIL), reaches 19.53 dB median PSNR and exceeds the strongest camera-adapted baseline by 0.79 dB on 18 of 18 targets (0.43 dB on 14 of 18 under a more faithful forward model). Most importantly, we show self-attention allows one shared network to serve 18 targets, and as the task complexity increases, the attention mechanism activates. These results establish self-attention as a scalable framework for far-field holography, and attention-based modeling as a general approach to inverse problems governed by long-range interactions in wave physics.

**Main Text:** Precise control of optical wavefronts underpins a broad class of photonic technologies, from holographic displays and adaptive optics to optical trapping, structured illumination microscopy, and free-space optical communications[1-9]. In each of these domains, the ability to generate arbitrary far-field intensity distributions from a phase-only modulator represents a central computational and physical challenge. A dominant approach to wavefront control is phase-only computer-generated holography (CGH), in which the optical field is shaped by encoding a phase pattern onto a spatial light modulator (SLM) [10,11]. The modulated wavefront then diffracts to reconstruct a desired intensity distribution in the target plane[12,13]. In practical systems, an SLM encodes a phase pattern onto an incident laser beam. The resulting wavefront is optically propagated through a Fourier lens, and the reconstructed intensity distribution is measured by a camera in the far field. This physical measurement loop enables direct evaluation of hologram quality and forms the basis of camera-in-the-loop (CITL) optimization commonly used in recent work[14-17]. Despite decades of progress, a scalable, non-iterative solution for phase-only hologram synthesis in the strict Fraunhofer regime, where propagation is fully global, has remained an open problem.

In the strict Fraunhofer regime, the reconstructed complex field is proportional to the Fourier transform of the SLM-plane field, so each SLM pixel contributes coherently to every point in the reconstructed field. This Fourier mapping is therefore fully dense, rendering phase-only hologram synthesis a non-local inverse problem in which local phase changes induce global intensity changes across the entire reconstructed field[12]. In practice, this global coupling imposes a persistent compute-fidelity trade-off at high resolution. It also amplifies sensitivity to forward model mismatch, since real SLMs exhibit spatially varying phase response and wavefront flatness that must be calibrated for accurate control, while additional optical aberrations and misalignment further distort the reconstructed field[6,15]. Furthermore, discrete phase levels and quantization alter diffraction efficiency and introduce structured error, especially when the available modulation range and phase linearity deviate from idealized assumptions[12,18]. Classical CGH algorithms solve this inverse problem via iterative phase retrieval[19]. The Gerchberg-Saxton (GS) algorithm and related Fourier-projection methods (iterative Fourier-transform algorithms, IFTA) alternate between enforcing constraints in the image and SLM planes[18,20]. While simple and widely used, these methods can stagnate in local minima and require many Fourier transforms to refine quality. More advanced optimization frameworks, including convex relaxations and gradient-based phase retrieval methods, have been proposed to improve convergence and robustness. However, these approaches typically rely on additional measurements or assume well-characterized forward models. They have seen limited success in phase-only holography under strict Fraunhofer propagation in practice, where the combination of dense Fourier coupling and hardware constraints remains challenging[21-24].

The far-field regime is not a niche alternative to Fresnel holography but a constraint imposed by a class of optical systems. Fresnel CGH holds real advantages: compactness, native three-dimensional light-field synthesis, and greater tolerance of the unmodulated zero order. Thus, near-eye display research has settled in the Fresnel regime for these reasons[25]. The case for Fraunhofer CGH is different. Wherever the working plane is the Fourier plane of the modulator by construction, the mapping from phase to pattern is the fully dense Fourier transform, and Fresnel formulation does not apply. This is the situation in optical trapping and structured-illumination microscopy, where the trap or illumination pattern lives at the objective's Fourier plane[6,26], in Fourier-plane beam shaping, and in adaptive-optics settings where the correction is applied in a pupil plane conjugate to the far field[3]. For these systems the choice is not Fraunhofer versus Fresnel, but Fraunhofer done well versus done poorly, and the methods available there are thin precisely because learned CGH has concentrated on the Fresnel regime.

More recently, deep learning has accelerated CGH synthesis, with neural networks trained to predict SLM phase from target images in a single forward pass[27-30]. The architectural choices underlying these advances are themselves a product of the propagation regime they target. Neural Holography[14] established CITL optimization for near-field holographic displays, in which holograms are displayed on the SLM and refined against measured reconstructions. But, its lineage: partially coherent extensions[16], dual-SLM systems[31], off-axis configurations[32], and learned hardware-in-the-loop phase retrieval[15], operates in the Fresnel regime, as do the Fourier-basis and Fourier-operator networks[33,34] and attention-augmented generators[35] of the recent learned-CGH literature. This concentration is not incidental. In Fresnel/angular spectrum method (ASM) settings (e.g. multi-depth or near-eye displays), the optical transfer retains partial spatial locality, which is exactly the structure that convolutional and local-window architectures exploit. Strict far-field propagation offers no such locality (Extended Data Fig. E10). It is a single Fourier transform of the full aperture, and every modulator pixel couples to every reconstruction point, so a conventional convolutional network would require an impractically large receptive field to capture the interactions. The CITL procedure itself is propagation-agnostic and transfers to the far field. The architectures and their inductive biases do not. The few far-field CITL studies bear this out. One uses the camera once, to calibrate a digital-twin forward model, after which it leaves the loop and hologram generation proceeds against the fitted model[17]. Another keeps the camera in the loop but pairs it with a convolutional network and optimizes an input-plane amplitude behind a fixed random phase, leaving the modulator phase itself outside the learned mapping[36]. The gradient-descent (GD) realization of per-iteration far-field camera feedback, used in this work as a baseline, shows promise for the far-field[37]. What has not been shown is an attention architecture aligned with the globally coupled nature of far-field propagation, performing hologram generation and experimental correction at once. A related lineage learns control of propagation through complex media directly from measured input-output pairs[38], in-situ learning of an unknown optical mapping that shares our motivation, though without the explicit Fourier structure exploited here. The classical Fourier-domain methods GS and GD remain the established baselines defined in the strict Fraunhofer regime, and they are the benchmarks used in this work. Consequently, a fully global, non-iterative network for far-field CGH, trained against the physical system it drives, has remained an open challenge[27,39].

Self-attention, originally developed for sequence modeling[40] and extended to vision through patch-based tokenization[41], enables direct modeling of long-range dependencies, allowing each element of a learned representation to interact with all others globally. This property mirrors, by construction, the structure of Fourier-optical propagation. Here we show that self-attention provides a natural framework for modeling the global coupling inherent to Fourier-optical wavefront control. We developed a self-attention-based hologram generator trained end-to-end through physical wave propagation, enabling direct generation of phase-only holograms from target intensity distributions in a single forward pass at inference. To account for system-specific deviations from the ideal model, we integrated self-attention-based camera-in-the-loop (SAIL) training, allowing the network to adapt to nonidealities in real optical systems. Together, this framework enables high-fidelity far-field hologram synthesis in a single forward pass, while capturing the global interactions and experimental imperfections that limit existing approaches. The advance, stated plainly, is a learned architecture whose parameterization matches the fully dense coupling of far-field propagation, trained directly against the physical optics it drives, with that adaptation shareable across targets. The network generates the hologram, no output of any other method enters the loop, and the optics sit inside training.

## Self-attention enables hologram generation in Fourier optics

The generator maps a target intensity to a phase hologram in one forward pass. The 1000 × 1000 target is divided into patches with patch size, p, of 500. They are half-field tokens, embedded to dimension 256, mixed by four multi-head self-attention encoder layers, and projected back to a complex field whose phase is the hologram (Fig. 1, steps 1-8; Algorithm 1 in Methods). Attention is the only stage where tokens interact. All experiments in this work are monochromatic (Methods).

To evaluate whether self-attention can capture the global coupling inherent to Fourier-optical propagation, we first assessed performance in simulation, where the forward model is known exactly and the ability of the network to represent the inverse mapping can be isolated. Results throughout are reported under two forward models. The ideal model treats every modulator pixel as a point sampler with unit fill, so the reconstruction is the plain Fourier transform of the phase field. The faithful model adds the sinc envelope of the finite pixel aperture and the device's measured areal fill factor of 0.91 with zero-padding so the resulting field is sampled without aliasing. It is the closer approximation to the physical system (Methods).

Two variants of the transformer are compared throughout. In the per-target transformer, a separate network is trained for each of the 18 targets, which isolates how well the architecture represents a single inverse mapping. In the batched transformer, one network is trained across all 18 targets at once and then serves any of them in a single forward pass, which tests whether the mapping is learned rather than fitted. Unqualified, the transformer means the per-target variant, and figures and tables label the two as Transformer and Batched Transformer. Both were trained for 10,000 epochs. The iterative baselines carry no weights, so they are run per target instead, at a matched operating point of 750 iterations, which is three times a typical published far-field phase-retrieval budget and, more to the point, the budget the CITL runs on the bench are given, so the simulation and hardware comparisons are matched, and separately swept to 10,000 iterations to locate where they converge (Methods). Training is paid once for the learned transformers but paid again for every target for the baselines, which is the distinction the compute comparison below turns on. Table 3e lists every method in this work, the budget it was given, and where each is reported.

At the operating point, the per-target transformer reached a median PSNR of 52.13 dB under the ideal model, against 37.39 dB for GS and 37.40 dB for GD (Fig. 2a). Normalized mean squared error (NMSE) follows the same ordering, roughly seven times below GD and fifty times below GS (Table 1). The margin is not carried by a favorable subset. Per target, the transformer exceeds GD on 18 of 18 targets under the ideal model and on 15 of 18 under the faithful one, though by a small margin (Fig. 2c). Under the faithful model the field compresses, with the per-target transformer at 15.0 dB, GD at 14.8 dB and GS at 9.7 dB (Fig. 2b). Unlike iterative approaches, which require repeated forward and inverse Fourier transforms, the transformer produced these holograms in a single forward pass at inference. Convergence behavior and wall-clock cost for both baselines are in Extended Data Fig. E5.

The batched transformer sits below both, at 27.7 dB median PSNR under the ideal model and 10.1 dB under the faithful one (Fig. 2a,b). This is expected, since one network of the same size must represent eighteen mappings rather than one on the same budget, so the shortfall is compute allocation, not architectural disability. Its value is amortization, and the hardware results below develop that.

We state the compute framing precisely because it is the claim the evidence supports. Neither method class converges at the matched budget, and given more compute, the picture inverts. Swept to 10,000 iterations under the ideal model, GD overtakes the per-target transformer, crossing it near 5,000 iterations and finishing 4.4 dB ahead at 56.49 dB median PSNR against 52.13 dB (Extended Data Fig. E5).

Neither the per-target nor the batched transformer was trained to convergence. Best checkpoints fell between epochs 9935 and 9999 of the 10,000-epoch budget with validation loss still falling (Methods), so the sweep sets GD at 10,000 iterations against a network that was still improving when its budget ended. We report the comparison in that direction because it is the one least favorable to the transformer. The advantage of the transformer is therefore not a superior converged optimum but compute at matched quality. Reaching 52.1 dB takes GD roughly 5,000 iterations and about 15 s per target, against 12 to 24 ms for a single forward pass (Table 2). Under the faithful model no such inversion occurs because every simulation-only method has already collapsed to within a few decibels of the others.

Attention is not the only architecture that couples the whole field, and the obvious alternative was tested. The Fourier neural operator (FNO)[42] learns its kernel directly in the frequency domain, which makes it the natural global candidate for a problem whose forward operator is a Fourier transform. We trained a standard, unmodified FNO per target under the ideal model at full mode coverage, 500 × 500 modes on the 1000 × 1000 grid, with 256.0 million parameters against the transformer's 195.7 million, for the same 10,000-epoch budget the transformer received. From random initialization it reached 8.75 dB median PSNR, the lowest of any method in this comparison, and the outcome was insensitive to learning rate across five orders of magnitude (Extended Data Fig. E1b). A second arm removed the inverse problem altogether. Rather than the far-field objective, the FNO was regressed onto GD's own converged phase, so the only question left was whether it can reproduce a hologram already known to work. It reached 37.43 dB, matching converged GD quality, so the architecture can represent a far-field hologram once it is shown one. What it cannot do is find one from the far-field objective as well under the same epoch count of 10,000. The gap is trainability, not capacity (Extended Data Fig. E1a, Table 1, Table 3d).

The token geometry was chosen by sweep, not by default. Holding embedding width, depth, head count and training schedule fixed, we varied the patch size, p, over 500, 250, 125 and 50, which takes the 1000 × 1000 field from four half-field tokens to 400 local ones. Median PSNR falls monotonically, from 53.22 dB at p = 500 to 36.19 dB at p = 250, 23.75 dB at p = 125 and 11.80 dB at p = 50, and the fall is monotone on 18 of 18 targets (Extended Data Fig. E2a). The reason is not that attention mixes less. Attention sees more tokens, not fewer, as p shrinks. The parameters live in the linear projections that map each token to and from its $2p^2$ pixel values. The output head alone, a single linear map from the 256-dimensional token code to $2p^2$ values, shrinks a hundredfold across the sweep, from 128 million parameters at p = 500 to 1.28 million at p = 50, while total parameters fall from 195.7 million to 5.2 million (Extended Data Fig. E2b). The sweep varies patch size with width, depth and schedule held fixed, so what it shows is that the head binds at this width and depth, not that finer tokens cannot work. It is also a simulation result under the ideal model, and this paper's own finding is that simulation ranking does not survive the optics. Finer patches under camera feedback are untested here and may yet prove useful on the bench, which we leave to future work.

The baselines are reported in their stronger configuration. GS and GD were run under both target formulations, amplitude and raw intensity, on all 18 targets, under the ideal (Extended Data Fig. E4a) and faithful (Extended Data Fig. E4b) forward models. For GD the two are equivalent. The implementation squares the amplitude back to intensity before the loss is formed, so both parameterizations optimize the same intensity-domain objective, and the paired differences are consistent with run-to-run variation, at +0.19 dB median PSNR under the ideal model (two-sided Wilcoxon signed-rank, $P = 0.28$) and +0.05 dB under the faithful one ($P = 0.20$), neither significant. For GS the formulation changes the algorithm because whichever array is chosen is substituted as the replay-plane magnitude constraint at every iteration, and substituting raw intensity distorts brightness quadratically. That costs 21.44 dB median PSNR under the ideal model on 17 of 18 targets and 0.45 dB under the faithful one on 16 of 18, both $P < 0.001$, with near-binary targets barely affected and continuous-tone scenes collapsing, which is the signature of that mechanism rather than of a tuning choice. The amplitude formulation used throughout is therefore the standard configuration for GS and an equivalent parameterization for GD.

That collapse from ideal to faithful representation is the result which matters most here. Simulation quality spans 27.7 to 52.1 dB median PSNR across GS, GD and the two transformers under the ideal model but only 9.7 to 15.0 dB under the faithful one, the first indication that fidelity of the forward model, not strength of the optimizer, is the binding constraint. The following section takes every one of these methods onto the bench and shows how far that constraint reaches.

### Bridging simulation and experiment with CITL

While the per-target transformer achieved strong performance in simulation, its advantage did not directly translate to experimental settings, nor did any of the other methods. We compared the transformer against classical iterative methods GS and GD under identical experimental conditions, by displaying the corresponding phase holograms on the SLM and capturing the resulting far-field intensity (Methods). As shown in Fig. 3, the transformer did not outperform these classical approaches when deployed on hardware, highlighting the impact of forward-model mismatch between simulation and the physical optical system. This indicated that self-attention alone does not translate simulation performance into experimental dominance. Under the ideal model every simulation-only method lands in the same narrow band on hardware: GS and GD at 750 iterations, GS and GD swept to 10,000, and the transformer in both its per-target and batched forms span only 15.8 to 16.1 dB median PSNR, despite spanning 27.7 to 56.5 dB in ideal simulation (Table 1).

Motivated by this gap, we introduce SAIL (self-attention in-the-loop), which places the physical optics inside the training loop. The transformer is optimized directly against camera measurements of its own reconstructions, so its weights absorb hardware-specific aberrations, misalignment and non-ideal phase response. Generation and adaptation are the same process in SAIL, instead of successive stages, which is what separates SAIL from CITL optimization, where the camera refines one target's phase field and nothing is learned that transfers to the next.

We evaluated the effectiveness of this approach by comparing the transformer alone to its SAIL counterpart. SAIL improved reconstruction fidelity on 17 of 18 targets, the exception falling short by 0.03 dB, yielding higher PSNR and lower NMSE in experiment (Fig. 3, Table 1). These gains confirmed that incorporating physical feedback is essential for translating learned hologram synthesis into real-world performance. More importantly, the network was able to adapt to the optical setup and alter phase to compensate for nonidealities. On hardware SAIL reaches 19.5 dB median PSNR against 16.1 dB for the simulation-trained transformer. Table 1 reports the mean and standard deviation alongside the median and interquartile range (IQR) for every method in both domains. The wide standard deviations in Table 1 reflect how much the 18 targets differ from one another, which was deliberate and does not represent instability in any method. Consistency is carried by the paired per-target differences in Fig. 3, where each point is a single target scored under both methods.

Qualitative results agree. The simulation-trained transformer's reconstructions carry speckle and lose fine structure, much like GS and GD, while SAIL recovers sharper edges and cleaner detail (Fig. 4, Extended Data Fig. E8). On the target shown in Fig. 4 the simulation-only methods land together between 16.99 and 17.08 dB, camera feedback lifts GD (GD+CITL in the plot) to 21.04 dB, and SAIL reaches 22.34 dB. SAIL generates the hologram at every stage and learns the system that generation is adapted to.

**Camera feedback does most of the work, and self-attention keeps its margin**

The comparison above leaves open whether SAIL's advantage comes from the attention architecture or simply from camera access. To separate the two, we gave the classical baselines direct access to the same camera measurements SAIL receives. We ran GS with CITL (GS+CITL) and GD+CITL on all 18 targets, under both forward models, in two initialization arms (random, and simulation-seeded from the converged simulation phase), and re-photographed every method in a single session on a single alignment (Fig. 3). GS+CITL was given 25 camera iterations rather than 750. Its best iteration is iteration zero on every target in every condition and the reconstruction had collapsed well before 25, identically on all 18, so further iterations would only have spent camera time confirming a failure already visible.

Each best hologram from each method was shown on the SLM and photographed in one session. Photographing every method within one session is what makes the comparisons below independent of rig drift. Re-photographing the same holograms in a separate session shifts scores by a median of 0.18 dB, with a per-target range of -1.01 to +0.05 dB, comparable to the rig's day-scale reproducibility of 0.14 dB (Methods). Both scorings are retained in the deposited record.

Camera feedback does most of the work. Per-iteration feedback in the random-initialization arm lifts GD by +2.18 dB median under the ideal model and +2.01 dB under the faithful one, over the simulation-only floor, confirming that adaptation to the physical system, not any property of a learned network, is the largest single factor on hardware. Self-attention then adds a further, consistent margin on top of identical feedback. SAIL is better than simulation-seeded GD+CITL, the strongest adapted baseline, by +0.79 dB median on 18 of 18 targets (+0.43 dB on 14 of 18 under the faithful model, both $P < 0.001$), and better than the random-initialization arm by +0.92 dB on 17 of 18 (+0.59 dB on 15 of 18). The comparison is stacked against us on purpose. The simulation-seeded arm is given 750 iterations of simulation optimization before the camera ever opens and 750 more with it, so SAIL at 750 epochs is measured against roughly twice the optimization budget of simulation-seeded GD+CITL and still leads on 18 of 18 targets. Both methods see identical camera information. The difference is what they do with it. A per-pixel phase update can only correct locally, while the attention layers propagate each measurement globally across the hologram.

Read across Fig. 3 as a whole, this is not a margin over one chosen baseline. SAIL is ahead of all eight comparators under both forward models: GS and GD at the 750-iteration operating point, the simulation-trained transformer in its per-target and batched forms, and both initialization arms of GS+CITL and GD+CITL. That is sixteen paired comparisons, and every one of them is positive. The margin runs from +0.43 dB median PSNR, against simulation-seeded GD+CITL under the faithful model, to +5.15 dB against random-initialization GS+CITL, and SAIL is better on 17 of 18 targets or more in 14 of the 16, never on fewer than 14. Every method was photographed in the same session on the same alignment, so the ordering is not assembled from separate campaigns.

GS+CITL fails outright. Its best iteration is iteration zero on every target in every condition. From random initialization it finishes below plain GS, by -1.06 dB under the ideal model and -0.75 dB under the faithful one, ahead on 1 of 18 targets in both. Seeded from the converged simulation phase it neither gains nor loses, at -0.01 dB and ahead on 8 of 18, which is what iteration zero means. It keeps its seed. The mechanism is structural. Alternating projection substitutes the full measured magnitude as its replay-plane constraint at every step, so measurement noise is injected directly into the constraint set and accumulates. CITL optimization is not a free upgrade that any iterative method absorbs.

A bias-dominance control addresses the architecture question from the other side. It asks whether the network computes its answer from the target at all, or has merely learned a fixed output, which is the cheapest explanation for a per-target result and has to be closed off before the selectivity evidence means anything. The diagnostic evaluates trained checkpoints through the simulated forward model, so it is reported in simulation. Removing the input-dependent pathway, so that only the learned bias reaches the output, collapses reconstruction from a median of 52.1 dB with the full network to 8.0 dB with the bias alone (27.7 to 8.0 dB batched), a median paired fall of 43.8 dB per target and 19.9 dB batched, on 18 of 18 targets in both regimes (Extended Data Fig. E3b). The network's quality lives in its input-dependent computation, not in a memorized offset.

**One network serves all targets**

The results above are per-target optimization, with a separate network trained for each image. To draw the distinction between per-target optimization and generalization explicitly, we trained one shared network jointly across all 18 targets in a single SAIL process (batched SAIL) and then served every target from the same weights, one forward pass each. At the 2,000-epoch budget, batched SAIL exceeds per-target SAIL by +0.15 dB median PSNR under the ideal model (better on 16 of 18 targets, $P = 0.004$) and is statistically at parity under the faithful model ($P = 0.23$). At 750 epochs it sits slightly below (-0.27 dB median PSNR). We report both budgets so that parity cannot be an artifact of stopping early (Fig. 5). The mapping from target to phase is being learned target-agnostically. The reason batched SAIL was extended to 2,000 epochs was to determine if the network itself would plateau at a lower quality per-target than SAIL which would be a limitation. As we have shown it, batched SAIL can achieve comparable performance if given enough compute, which is expected.

The parity is a hardware result. In ideal simulation the batched transformer lags its per-target counterparts by a wide margin (27.66 against 52.13 dB median PSNR), yet on the bench batched SAIL matches per-target SAIL. Batched SAIL's parity emerges under hardware adaptation.

Batched training is not cheaper than per-target training. Both spend one camera exposure per target per epoch, and the batched sessions cost 8.35 hours (750 epochs) and 23.03 hours (2,000 epochs) against total 6.82 hours for the 18 per-target runs under the ideal model (Table 2), all on a single NVIDIA RTX 4000 Ada Generation laptop GPU with 12 GB of memory (Methods). What batching changes is what the spend buys. It yields one network instead of eighteen, which drops the marginal cost of serving a new target from a full SAIL run, roughly 23 minutes, to a single forward pass of tens of milliseconds. That is the practical face of the generalization claim. That amortization is structurally unavailable to the classical baselines and to many of the modern techniques in hologram generation, especially in the far-field. GS and GD have no shared weights, so no operation batches 18 independent phase-retrieval problems into one training process.

Now, whether the mapping itself transfers is a separate question, and a separate study answers it. A small network trained at 28 × 28 on a narrow curriculum, three classes of MNIST handwritten digits[43], 2, 6 and 9, together with synthetic patterns, was then asked for targets it had never seen: the other seven digit classes, geometric shapes and free-form drawings. It produced recognizable reconstructions for all of them (Extended Data Fig. E6, Methods). The mapping from target to phase is learned, not fitted one picture at a time. The low resolution is a training-budget choice rather than a limit of the architecture, since the main-text network already runs at the full 1000 × 1000 grid.

The attention maps show where the extra work happens, and they answer a question the per-target results cannot. Per-target synthesis is a fitting problem. GD is nothing but a free per-pixel phase field, and it reaches 37.4 dB at the operating point and 56.5 dB when swept to 10,000 iterations, so success on one target shows only that the parameterization has enough freedom. It cannot show that any particular architecture is contributing. The attention maps say the same thing from the inside. We measure selectivity as $1 - H/\log T$. Here $H$ is the Shannon entropy of one head's attention distribution over the $T$ tokens, the standard ($-\Sigma\, a \log a$) over the weights $a$ a head assigns, and $\log T$ is the largest value that entropy can take, reached when a head spreads its weight equally over every token. Dividing by $\log T$ removes the dependence on how many tokens there are, so the measure is comparable across configurations, and subtracting from one turns it into a deficit, meaning how far a head has moved away from mixing everything equally and toward committing to a few tokens. Zero is uniform mixing, one is a head reading a single token. Training one network per target leaves selectivity at 0.049 in simulation and 0.078 on hardware. The measurement covers four training regimes, per-target and batched, each trained in simulation and through the physical optics, and averages over every head in every layer on all 18 targets. The attention maps it is computed from are shown for all four regimes in Extended Data Fig. E3a. Near-uniform is what should be expected there, which is why the comparison is informative rather than a null result. Attention exists to make a computation depend on its input, to route different information for different inputs. A network trained on one target only ever sees one input, so there is nothing to route conditionally. Any fixed pattern of mixing suffices because the projections downstream can absorb whatever the attention hands them, and the answer they must produce never changes. Nothing in the objective rewards a head for preferring one token over another, so the heads stay close to where initialization left them, which is close to uniform. The heads are present and very nearly idle because memorizing one mapping does not require them.

Requiring one network to serve all 18 targets changes exactly that. The same weights must now produce a different hologram for each input, and attention is the only stage in this architecture where the computation can depend on which target it was given, so commitment stops being optional. Selectivity rises to 0.525 in simulation and 0.654 through the physical optics, an order of magnitude above the per-target values (Fig. 5c, with the corresponding attention maps for all four regimes in Extended Data Fig. E3a). Architecture, token geometry, parameter count, and data format are identical across all four conditions. Only the demand on the network changed. Per-target training stays near uniform on every one of the 18 targets and the batched networks are selective on every one (Extended Data Fig. E3c). The rise is not a training-length artifact either. The batched network trained through the optics for 750 epochs reaches 0.712, marginally above the 0.654 of the 2,000-epoch network quoted above, so selectivity is not something that accumulates with a longer budget. Real optics ask more of the mechanism than the simulated model does, in both regimes. Attention is not decoration that happens to be present when the network works. It is recruited exactly when the task stops being fittable per target and starts demanding a mapping that generalizes.

### SAIL adaptation is specific to the optical state it is trained on

Optical misalignment introduces phase distortions and calibration errors that degrade reconstruction quality[44,45]. To characterize how SAIL training behaves under such perturbations, we displaced the Fourier lens toward the camera by nominal caliper settings of 0.05, 0.1, 1 and 2 mm (uncertainty approximately ±0.05 mm), on a single target with one frozen calibration and a fixed exposure across the sweep (Methods). Three arms were compared at each position: the never-adapted, transformer-generated simulation hologram, a SAIL hologram trained on the aligned bench, and SAIL retrained at that displacement.

Adaptation is specific to the optical state it observes. A hologram trained at 2 mm and displayed back on the aligned system scores 7.00 dB against 7.61 dB for the hologram trained at zero, a loss of 0.61 dB and roughly fifteen times the 0.04 dB measurement floor of this experiment (Extended Data Fig. E9). Return-to-zero checks between conditions scored within +0.03 dB for SAIL and +0.04 dB for the simulation-only arm, both far inside the rig's day-scale reproducibility of 0.14 dB, so the effect is not drift.

That specificity does not amount to correction. At 2 mm displacement the three arms converge, at 7.16, 7.17 and 7.27 dB PSNR, so retraining in place buys 0.11 dB over never adapting at all and the reconstructions are visually indistinguishable (Fig. 6). Across the sweep, retrained SAIL never returns to baseline, falling 0.07, 0.06, 0.18 and 0.38 dB below it as displacement grows, and the SSIM advantage of the aligned-bench hologram over the never-adapted arm decays from 0.059 to 0.002 (Extended Data Fig. E9). SAIL training locks onto the state of the optics it sees, but it cannot restore quality that defocus has physically removed. Where good quality is achievable, being adapted to the right state matters a great deal. Where the optics have thrown quality away, adaptation has little left to exploit.

### Implications and outlook

Attention establishes itself as a practical operating mode for generating holograms for Fraunhofer (far-field) photonic systems, and SAIL is how that architecture is fitted to a physical one. A single network generates the hologram in one forward pass, and because the optics sit inside the training loop, that generation is adapted to the physical hardware it drives rather than to an idealized model of it. While the results presented here focus on per-target optimization, a shared network trained across multiple targets achieves performance comparable to individually optimized networks (Fig. 5), indicating that the mapping from target intensity to phase modulation can be learned in a target-agnostic manner. This enables substantially increased throughput, as a single network can generate phase holograms for diverse targets in real time without retraining.

Consistent with this, the network captures transferable structure in the Fourier domain instead of memorizing individual examples, which is what allows single-shot prediction of phase holograms for targets never seen in training, with no per-target iterative optimization[46,47] (Extended Data Fig. E6). Scaling that demonstration from the 28 × 28 proof of concept to the full grid is a matter of training budget and of assembling a curriculum that spans the full frequency content, and is left to future work. It is also where the argument of this paper points. Attention is what makes a single network able to serve targets it was never trained on, and that, rather than any single reconstruction, is what would make far-field holography practical at scale.

These properties are particularly relevant for applications where global optical coupling and experimental nonidealities limit performance. In adaptive optics, learned wavefront control could enable rapid aberration correction without iterative optimization, within the bounds the aberration study establishes. Adaptation is specific to the optical state under which it was trained, and it does not restore quality that the optics have physically removed (Extended Data Fig. E9). In holographic displays, single-shot inference combined with experimental adaptation may support high-resolution, real time image generation under hardware constraints. Similarly, in optical trapping and microscopy, the ability to generate and refine complex light fields directly from target patterns may enable more flexible and robust control of structured illumination.

One practical consideration for deployment is the unmodulated zero-order component, which this work excludes from scoring by a fixed mask, identical for every method (Methods), and does not mitigate. The standard optical mitigation routes apply unchanged because none of them interacts with the learning loop. An off-axis carrier displaces the working reconstruction away from the zero order, a physical beam block at the Fourier plane removes it at the cost of the DC region, and complementary-frame or polarization multiplexing suppresses it temporally[48]. SAIL is compatible with each, since they modify the target geometry or the optical train rather than the generation algorithm. SAIL would simply see, and adapt to, the modified system. We claim no measured performance under any of them.

More broadly, recent work has highlighted the potential of combining physical systems with learned models to directly control complex optical transformations in situ[26,38,49]. Our results extend this direction to the fully non-local regime of Fourier optics, where global coupling and hardware mismatch have historically limited both classical and learned approaches.

Together, these results point toward a framework in which learned models and physical feedback are jointly leveraged to achieve high-fidelity wavefront control under realistic conditions. This work suggests a shift from purely physics-driven inverse solvers toward hybrid, experimentally grounded learning-based approaches for controlling complex optical wavefronts, enabling high-fidelity hologram generation from single-shot measurements without temporal or ensemble averaging and supporting real-time operation.

### Limitations

The evidence in this work comes from one optical system: a single spatial light modulator, a single wavelength, one Fourier lens and one alignment, scored on 18 targets. Nothing in SAIL is tied to that hardware, but nothing here demonstrates transfer to another bench either, and the adaptation SAIL learns is by construction specific to the optical state it was trained on. A hologram trained at one lens position loses 0.61 dB when displayed at another, and retraining under a displaced lens does not recover what the defocus has physically removed. SAIL also requires camera access during training and a training session per optical system, which is a real cost that the single forward pass at inference does not remove. Operation is monochromatic, and color would need wavelength-multiplexed illumination the present system does not have.

Several results are bounded by the conditions under which they were measured. The patch-size sweep is simulation-only and under the ideal forward model, so it shows that the output head binds at this width and depth, not that finer tokens fail on the bench, where camera feedback may change the answer. Generalization to targets never seen in training is demonstrated at 28 × 28 with a 3.3-million-parameter network, not at the full grid. Scaling it is a matter of training budget and of assembling a curriculum that spans the frequency content, and remains future work. That it works at 28 × 28 is an indication, not a guarantee that it holds at full modulator resolution. The FNO baseline is run under the ideal model only and its variant was selected in simulation rather than under camera feedback. The unmodulated zero order is excluded from scoring by a fixed mask applied identically to every method, and is not mitigated. Neither the learned nor the iterative methods were run to convergence at the matched budget, and given roughly thirteen times the compute gradient descent overtakes the transformer in ideal simulation, which is the comparison we report because it is the one least favorable to us. All comparisons are over the same 18 targets with no correction for multiple comparisons, and the two forward models are separate scoring problems that are never merged.

The most consequential limitation is absolute quality. SAIL reconstructs better than every comparator we ran, but better than the comparators is not the same as good enough for what far-field applications demand, and we have not characterized the network against the requirements of any one of them. Where that threshold sits for adaptive optics, for optical trapping, for display, and whether SAIL clears it, is open. What these results do establish is that attention trained inside the optics it drives is a route worth taking to the problems Fourier optics poses.

## Conclusion

In summary, we show that self-attention provides a natural framework for modeling the global coupling inherent to Fourier-optical wavefront control. While learned models alone achieve strong performance under idealized conditions in simulation, experimental results reveal that physical adaptation remains essential for high-fidelity reconstruction. By training that architecture inside the optics it drives, we establish a unified approach that in practice enables accurate and robust hologram synthesis in real optical systems, and that generalizes: one network serves targets it never saw in training. Beyond improving reconstruction quality, this work reframes phase-only holography as a learnable, globally coupled mapping that can be efficiently approximated and experimentally refined via self-attention. Beyond holography, this work suggests a broader principle. Aligning learned model architecture with the physical symmetries of a system, instead imposing local inductive biases, may unlock scalable, non-iterative solutions to inverse problems across photonics and wave physics more generally.

**Data availability:** The complete experimental and simulation record, including camera captures, run logs, configuration snapshots, convergence traces and scored results, is deposited in the University of Cambridge Apollo repository at https://doi.org/10.17863/CAM.133088, under embargo until publication. The GD-arm files of the CITL baseline runs are in the related deposit at https://doi.org/10.17863/CAM.132918. The eighteen target images are third-party material, and their sources and licenses are listed in the deposit.

**Code availability:** All method and analysis code is available at https://github.com/dilawersingh/sail-far-field and is included in the Apollo deposit at https://doi.org/10.17863/CAM.133088. A verification script confirms the deposit is complete and intact, and the deposited analysis notebook regenerates every figure, table and reported number from the deposited record alone. A hardware-free demonstration trains the generator against the simulated forward model on any machine. Trained model weights exceed the repository's upload limits and are available from the corresponding author on request.

**Acknowledgements:** The authors thank the University of Cambridge and the Department of Engineering (Electrical Engineering Division) for access to the resources and facilities that made this work possible. The authors also thank the Cambridge Trust and EPSRC for financial support.

**Funding:** This work was supported in part by the Cambridge Trust and EPSRC Centre for Doctoral Training in Connected Electronic and Photonic Systems [grant number EP/S022139/1].

**Author contributions:** Dilawer Singh conceived the idea and study, developed the methodology, performed all simulations and experiments, analyzed the data, and wrote the manuscript. Antoni Wójcik contributed to the development of the methodology through sustained technical discussion throughout the project, over numerous lunches and uncountable coffees. Professor Timothy Wilkinson supervised the project and provided overall guidance. All authors reviewed and contributed to the final manuscript.

**Competing interests:** The authors declare no competing interests.

Supplementary information is available for this paper.

Correspondence and requests for materials should be addressed to Dilawer Singh (ds2070@cam.ac.uk)

## Figure legends:

**Figure 1 | SAIL, self-attention in-the-loop hologram generation.** The method as a numbered walkthrough (steps 1-13). The 1000 × 1000 target intensity (1) is split into patches using p = 500 into four half-field tokens (2). Tokens are flattened and linearly embedded to dimension 256 (3), given a learnable 2D positional embedding over the 2 × 2 token grid (4), and processed by four transformer encoder layers (16 heads, feedforward dimension 1024, pre-norm), the only stage where tokens exchange information (5). A linear head returns $2p^2$ values per token in two channels (6), reverse-patched to the complex field Y (7). Phi = arg(Y) is the phase hologram (8). The network generates the hologram. No hologram from any other method enters this loop. The walkthrough forks at the hologram. On the physical branch (9a), phi is quantized to 8 bits and displayed on the SLM (Holoeye LUNA), the field propagates through the physical Fourier optics (f = 100 mm) to the camera (10), one exposure per epoch, calibrated (ROI crop, de-rotation, resampling, floor subtraction, normalization) to give I_cam (11). On the simulated branch (9b), the same field passes through the differentiable Fraunhofer model to give I_sim. The branches meet at the straight-through junction (12). I_proxy = I_sim + stopgrad(I_cam - I_sim), so the camera supplies the value and the simulated model supplies the gradient. The loss is L = L_cam + 0.1 L_sim, with the zero order excluded from both terms. AdamW then updates the network weights, steps 3-7 (13). The reported hologram is the checkpoint with the lowest camera loss. 195,660,320 parameters and one forward pass per hologram. Steps are keyed one-to-one to Algorithm 1 (Methods).

**Figure 2 | Simulation performance under both forward models.** PSNR over the 18 targets, with the iterative baselines at their 750-iteration operating point, the budget the CITL runs also use. Their converged behavior is in Extended Data Fig. E5 and Table 1. **a**, Ideal forward model. Each point is one target and the bar marks the median with the interquartile range. The per-target transformer reaches 52.1 dB against 37.4 dB for both GS and GD, and the batched transformer reaches 27.7 dB. **b**, The same comparison under the faithful forward model, where the methods compress to within a few dB, GD at 14.8 dB and the per-target transformer at 15.0 dB, with GS at 9.7 dB and the batched transformer at 10.1 dB. **c**, Per-target parity against GD at the operating point. Each marker is one target, plotting the transformer's single-pass PSNR against GD at 750 iterations, with the dashed line marking parity. The transformer is above parity on 18 of 18 targets under the ideal model and 15 of 18 under the faithful one by a small margin.

**Figure 3 | SAIL against every alternative on the bench.** Paired per-target difference in PSNR, SAIL minus each comparator, over the 18 targets. ***a***, Ideal forward model. ***b***, Faithful forward model. Each faint marker is one target, the filled circle is the median of the 18 paired differences and the bar spans its 95% bootstrap confidence interval, with that median and the number of targets favoring SAIL annotated at the right of each row. Positive values favor SAIL. The eight comparators are GS and GD at the 750-iteration operating point, the simulation-trained transformer in its per-target and batched forms, and GS+CITL and GD+CITL in both initialization arms, random and simulation-seeded. Every method was photographed in a single capture session on a single alignment. All sixteen comparisons are positive, from +0.43 dB against simulation-seeded GD+CITL under the faithful model to +5.15 dB against random-initialization GS+CITL. Absolute values for every method, mean and standard deviation beside median and interquartile range, are in Table 1, and the convergence of the iterative baselines beyond 750 iterations is in Extended Data Fig. E5.

**Figure 4 | SAIL against every method on one target, full frame and close up.** Bench reconstructions of one target, the alley scene, photographed in the single-session capture on one alignment under the ideal forward model, with canonical full-frame PSNR printed on each

panel. SAIL leads its strongest adapted comparator, simulation-seeded GD+CITL, by +1.30 dB on this target. ***a***, Full-frame reconstructions for the target, GS, GD, the simulation-trained transformer, GD+CITL, SAIL and batched SAIL, on a shared display scale. ***b***, The same methods over an enlarged region of interest, with the region marked on the target. The simulation-only methods land together between 16.99 and 17.08 dB and are visibly speckled, camera feedback lifts GD+CITL to 21.04 dB, and SAIL reaches 22.34 dB with batched SAIL at 22.76 dB. The GD+CITL arm shown is the simulation-seeded one and Batched SAIL is the 2,000-epoch budget. Region coordinates are versioned with the released code, and panels are reproduced at the full resolution of the analysis output. Extended Data Fig. E8 shows five targets spanning the range of that margin under both forward models, to show the range of performance rather than a single case.

**Figure 5 | One network serves all targets.** Batched SAIL trains one shared network across all 18 targets in a single SAIL process and serves any target in one forward pass. ***a***, Bench parity with per-target SAIL at both training budgets, with +0.15 dB median over per-target SAIL at 2,000 epochs under the ideal model (better on 16 of 18 targets, P = 0.004), statistical parity under the faithful model (P = 0.23), and -0.27 dB median at 750 epochs. ***b***, Marginal cost of serving one target, on a logarithmic axis spanning five orders of magnitude. Batched training is not cheaper, since both spend one camera exposure per target per epoch (Table 2), but it yields one network instead of eighteen, dropping the cost of serving a new target from a full SAIL run, roughly 23 minutes, to a single forward pass. ***c***, Attention selectivity rises with task complexity. Selectivity is near zero when one network serves one target in simulation, and rises by an order of magnitude when one network must serve all 18 through real optics.

**Figure 6 | Adaptation under controlled misalignment.** The Fourier lens displaced 2 mm toward the camera, one target, one frozen calibration, fixed exposure. The target and an enlarged region of interest are shown at left, then single-capture reconstructions from three arms, the never-adapted, simulation-transformer hologram, a SAIL hologram trained on the aligned bench when the lens was at 0.0 mm, and SAIL retrained at the 2 mm displacement. The three land within 0.11 dB of one another against a 0.04 dB measurement floor, so in-loop adaptation recovers little of what defocus has removed. The displacement sweep and the converse control are in Extended Data Fig. E9.

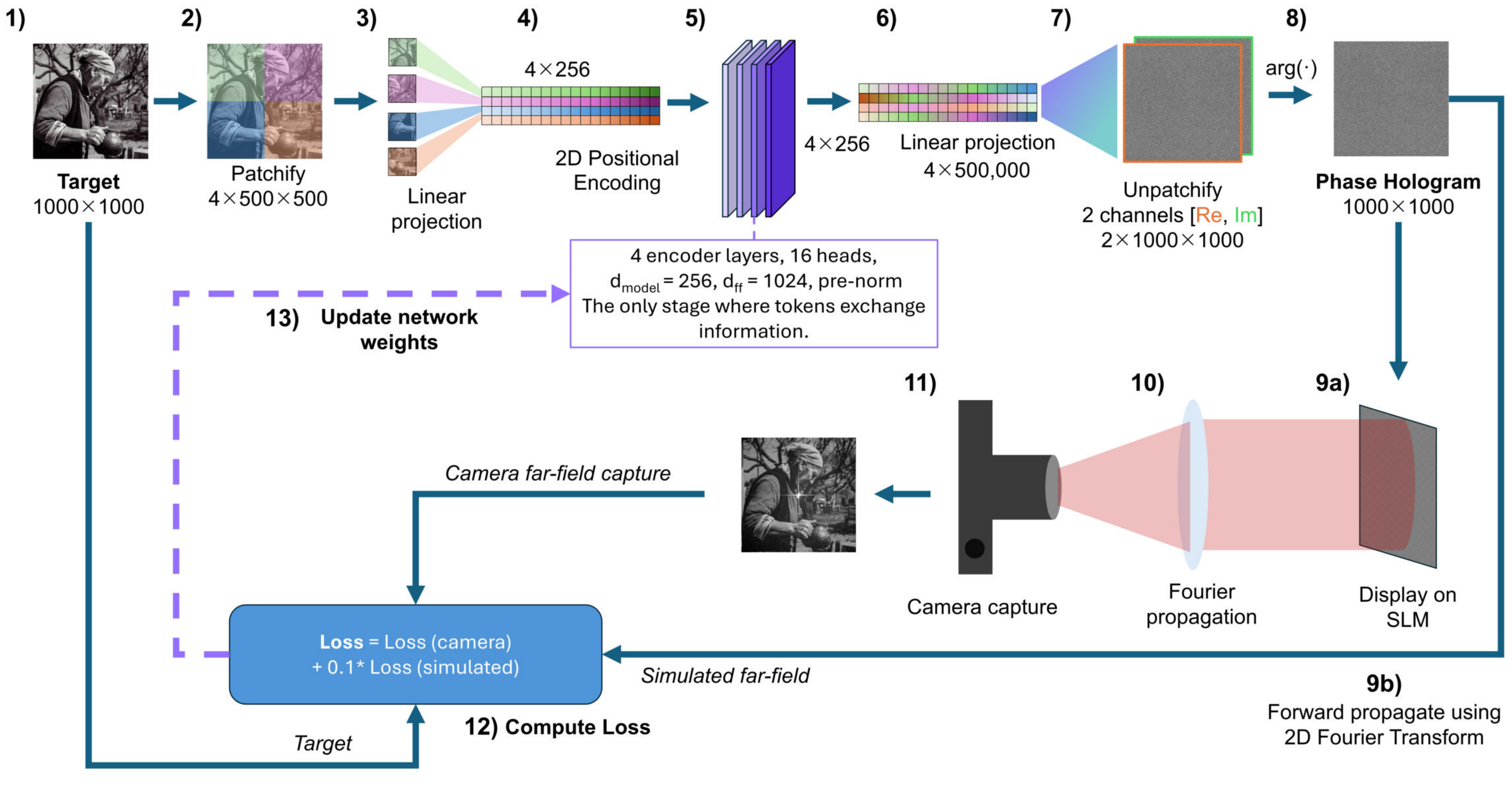
1)
Target
1000×1000
2)
Patchify
4×500×500
3)
Linear projection
4)
4×256
2D Positional Encoding
5)
4 encoder layers, 16 heads,
d_model = 256, d_ff = 1024, pre-norm
The only stage where tokens exchange information.
6)
4×256
Linear projection
4×500,000
7)
Unpatchify
2 channels [Re, Im]
2×1000×1000
8)
arg(·)
Phase Hologram
1000×1000
9a)
Display on SLM
9b)
Forward propagate using 2D Fourier Transform
10)
Fourier propagation
11)
Camera capture
Camera far-field capture
Simulated far-field
Loss = Loss (camera)
+ 0.1* Loss (simulated)
12) Compute Loss
Target
13) Update network weights

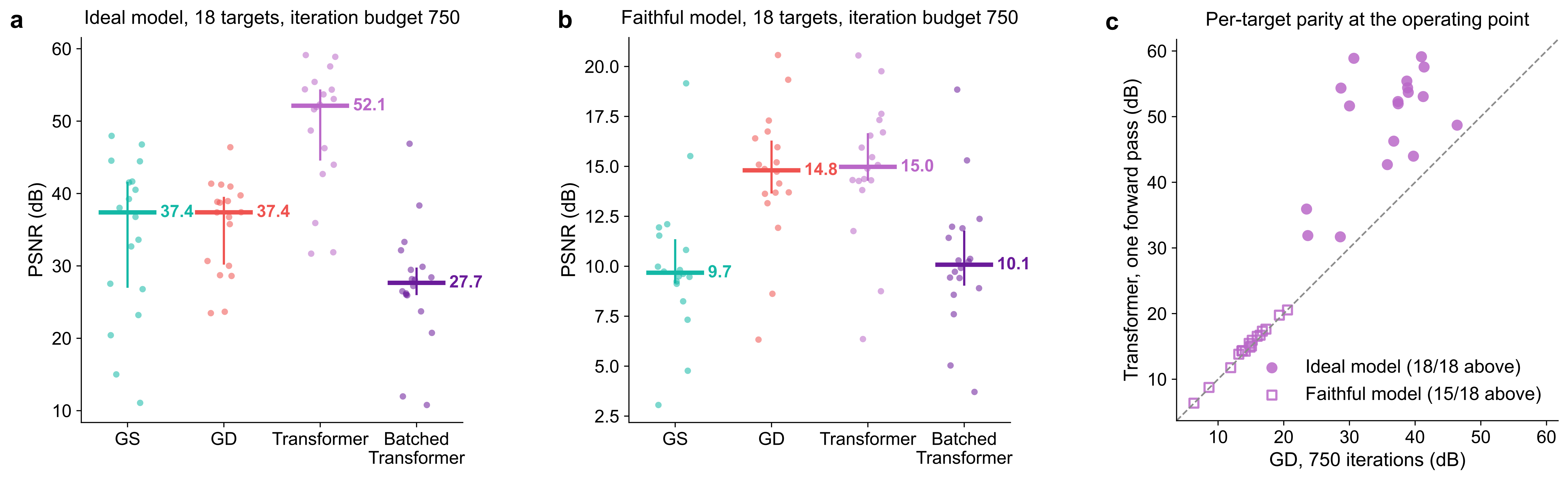
a
Ideal model, 18 targets, iteration budget 750
PSNR (dB)
37.4
37.4
52.1
27.7
GS
GD
Transformer
Batched Transformer
b
Faithful model, 18 targets, iteration budget 750
PSNR (dB)
9.7
14.8
15.0
10.1
GS
GD
Transformer
Batched Transformer
c
Per-target parity at the operating point
Transformer, one forward pass (dB)
GD, 750 iterations (dB)
Ideal model (18/18 above)
Faithful model (15/18 above)

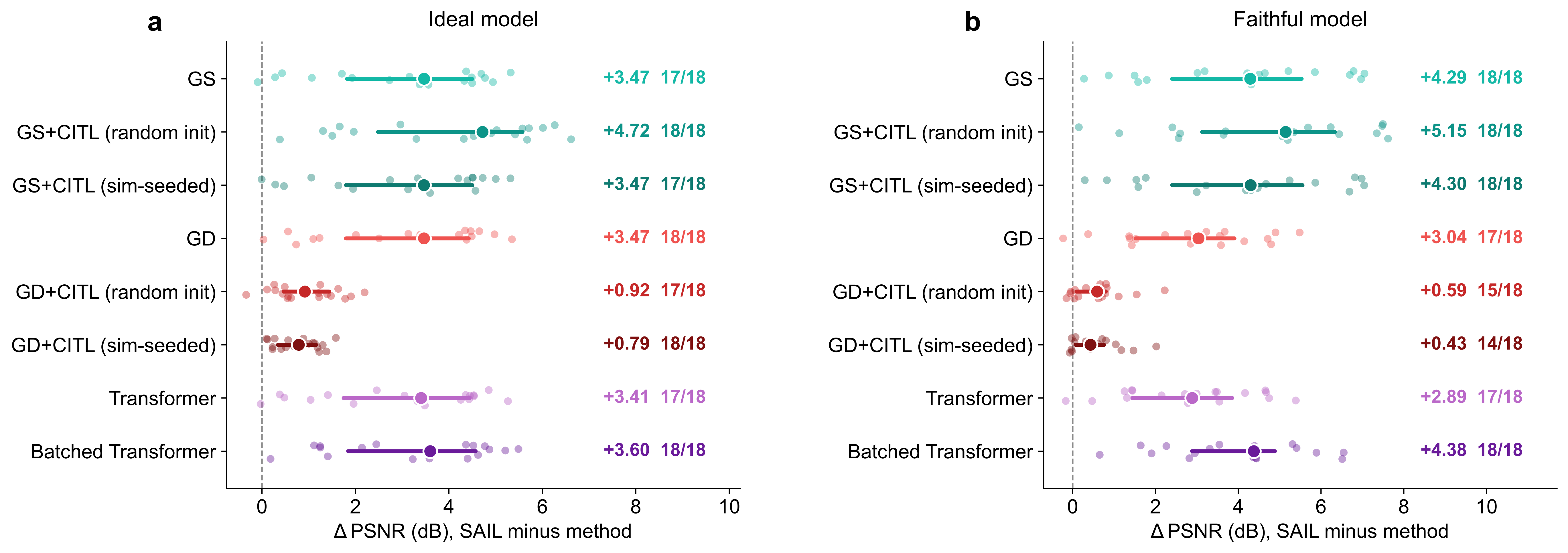

a
Ideal model
GS
GS+CITL (random init)
GS+CITL (sim-seeded)
GD
GD+CITL (random init)
GD+CITL (sim-seeded)
Transformer
Batched Transformer
+3.47 17/18
+4.72 18/18
+3.47 17/18
+3.47 18/18
+0.92 17/18
+0.79 18/18
+3.41 17/18
+3.60 18/18
0
2
4
6
8
10
Δ PSNR (dB), SAIL minus method
b
Faithful model
GS
GS+CITL (random init)
GS+CITL (sim-seeded)
GD
GD+CITL (random init)
GD+CITL (sim-seeded)
Transformer
Batched Transformer
+4.29 18/18
+5.15 18/18
+4.30 18/18
+3.04 17/18
+0.59 15/18
+0.43 14/18
+2.89 17/18
+4.38 18/18
0
2
4
6
8
10
Δ PSNR (dB), SAIL minus method

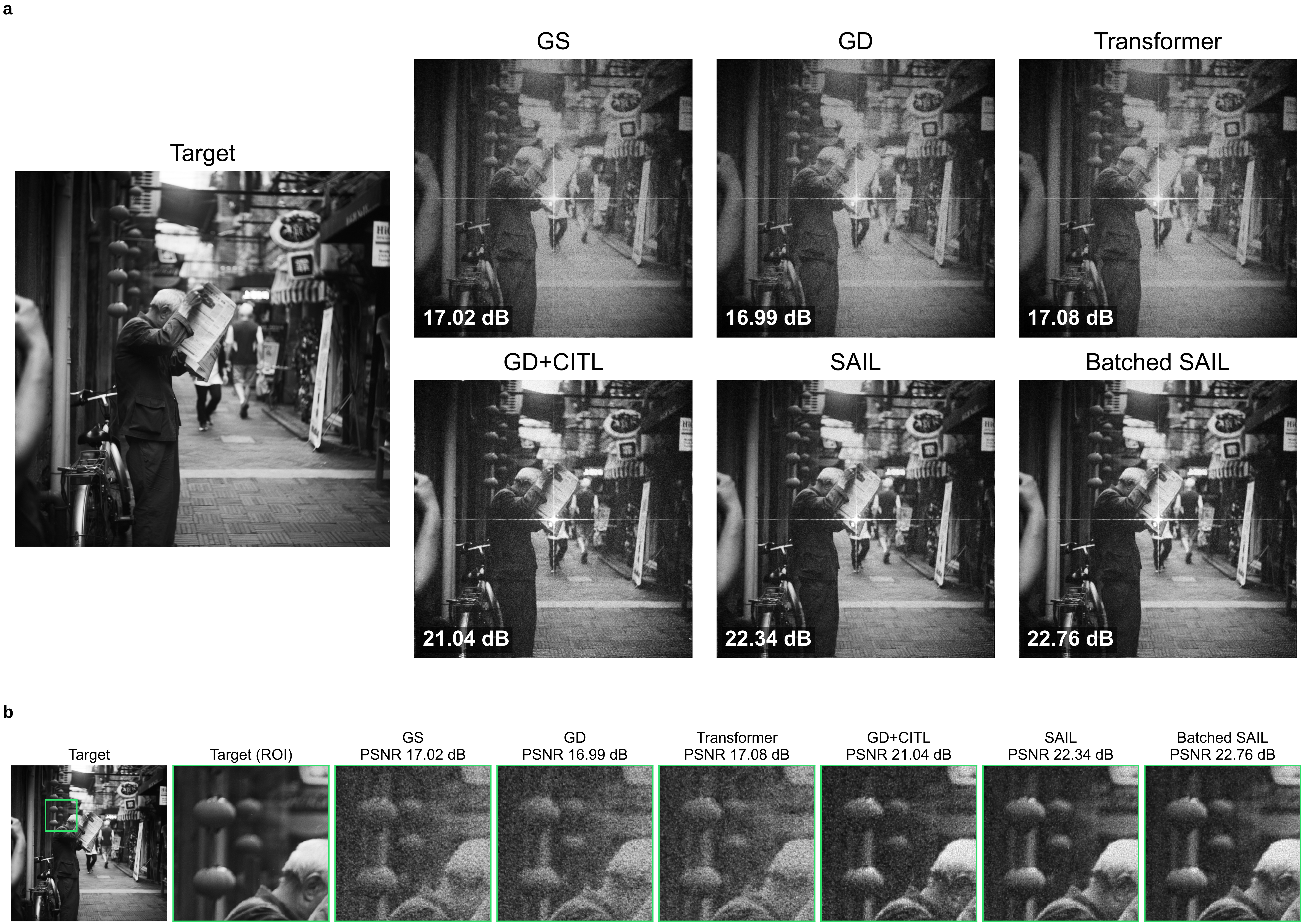

a
Target
GS
GD
Transformer
17.02 dB
16.99 dB
17.08 dB
GD+CITL
SAIL
Batched SAIL
21.04 dB
22.34 dB
22.76 dB
b
Target
Target (ROI)
GS
PSNR 17.02 dB
GD
PSNR 16.99 dB
Transformer
PSNR 17.08 dB
GD+CITL
PSNR 21.04 dB
SAIL
PSNR 22.34 dB
Batched SAIL
PSNR 22.76 dB

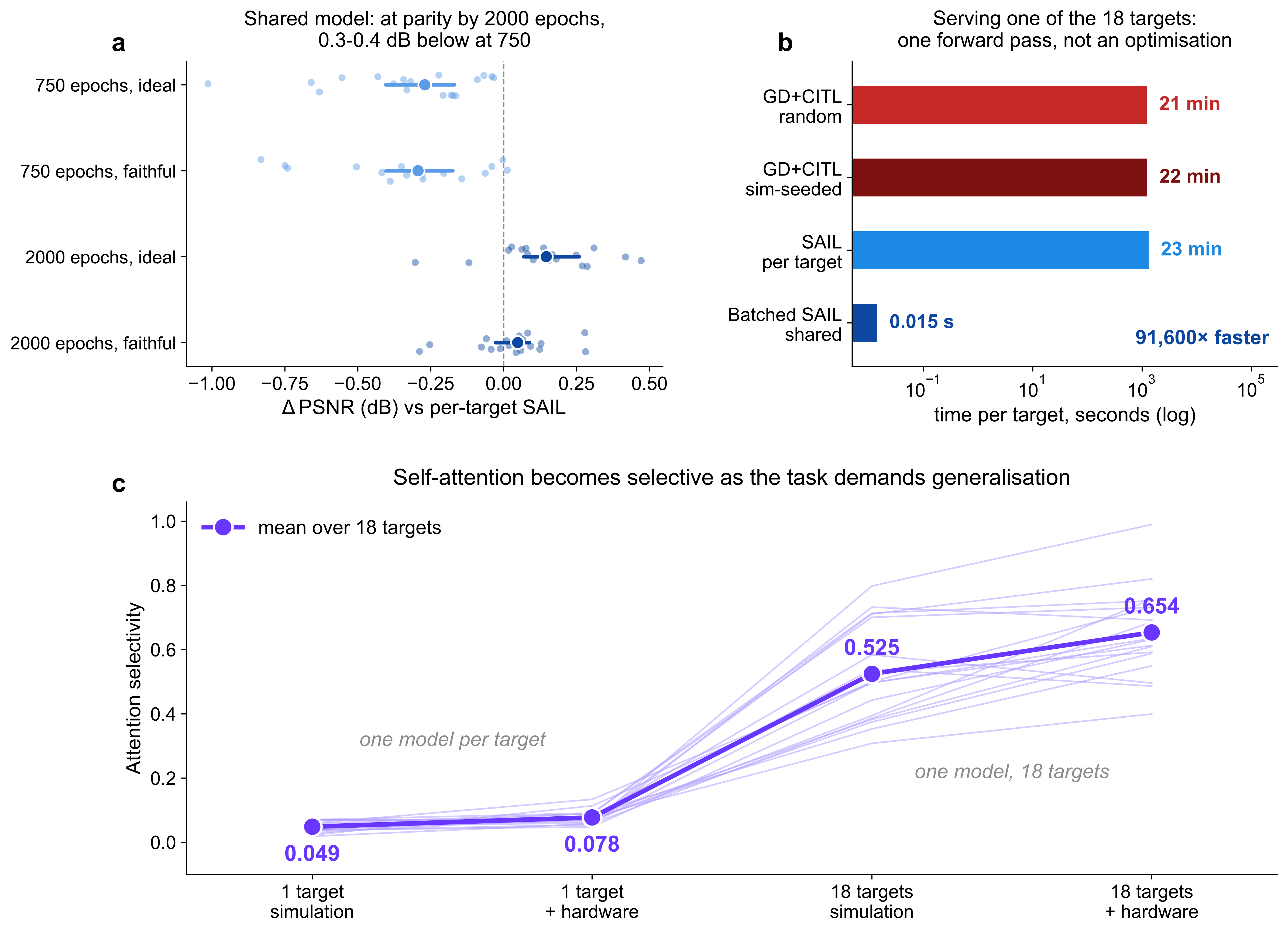

a
Shared model: at parity by 2000 epochs, 0.3-0.4 dB below at 750
750 epochs, ideal
750 epochs, faithful
2000 epochs, ideal
2000 epochs, faithful
−1.00
−0.75
−0.50
−0.25
0.00
0.25
0.50
Δ PSNR (dB) vs per-target SAIL
b
Serving one of the 18 targets: one forward pass, not an optimisation
GD+CITL random
21 min
GD+CITL sim-seeded
22 min
SAIL per target
23 min
Batched SAIL shared
0.015 s
91,600× faster
time per target, seconds (log)
c
Self-attention becomes selective as the task demands generalisation
mean over 18 targets
Attention selectivity
one model per target
one model, 18 targets
0.049
0.078
0.525
0.654
1 target simulation
1 target + hardware
18 targets simulation
18 targets + hardware

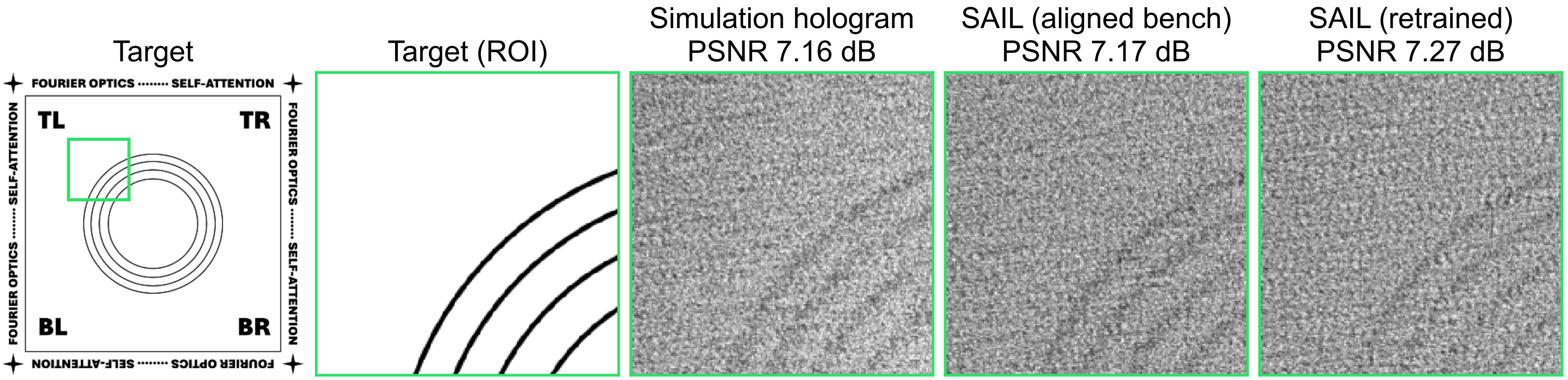
Target
Target (ROI)
Simulation hologram
PSNR 7.16 dB
SAIL (aligned bench)
PSNR 7.17 dB
SAIL (retrained)
PSNR 7.27 dB
FOURIER OPTICS ········ SELF-ATTENTION
TL
TR
BL
BR

**Tables:**

**Table 1 | Reconstruction quality for every method.** Mean ± SD beside median and interquartile range (IQR) over the 18 targets, with SSIM and NMSE (mean ± SD), under both forward models. (a) Simulation. (b) Bench, with every method photographed in a single capture session on a single alignment. The narrative quotes medians. Means and SDs are reported here for every method. The SD is spread across 18 deliberately heterogeneous targets (between-target heterogeneity). The paired per-target statistics carry the consistency of each comparison. The intensity-target rows are the target-formulation control of Extended Data Fig. E4, and FNO is reported under the ideal forward model only (Methods). Tables and figures label the simulation-seeded initialization arm as sim-seeded.

**(a) Simulation.**

| Model | Method | PSNR (dB) mean ± SD | PSNR median [IQR] | SSIM mean ± SD | NMSE mean ± SD |
|---|---|---|---|---|---|
| Ideal model | GS | 33.99 ± 11.03 | 37.39 [26.98, 41.62] | 0.911 ± 0.142 | 0.0268 ± 0.0506 |
| | GS (10,000 iterations) | 42.72 ± 17.07 | 45.95 [27.85, 58.81] | 0.933 ± 0.133 | 0.0228 ± 0.0499 |
| | GS (intensity target) | 16.61 ± 3.46 | 15.37 [14.74, 17.39] | 0.590 ± 0.141 | 0.1653 ± 0.0527 |
| | GD | 35.48 ± 6.45 | 37.40 [30.18, 39.54] | 0.851 ± 0.162 | 0.0034 ± 0.0037 |
| | GD (10,000 iterations) | 53.23 ± 12.30 | 56.49 [44.22, 63.82] | 0.940 ± 0.135 | 0.0006 ± 0.0010 |
| | GD (intensity target) | 35.43 ± 6.23 | 37.21 [30.46, 39.66] | 0.853 ± 0.162 | 0.0032 ± 0.0031 |
| | Transformer | 49.08 ± 8.65 | 52.13 [44.55, 54.38] | 0.933 ± 0.139 | 0.0005 ± 0.0009 |
| | Batched Transformer | 27.43 ± 8.22 | 27.66 [25.96, 29.77] | 0.622 ± 0.198 | 0.0239 ± 0.0340 |
| | FNO (from scratch) | 10.48 ± 8.16 | 8.75 [7.90, 10.76] | 0.221 ± 0.192 | 0.7394 ± 0.2474 |
| | FNO (regression onto GD phase) | 35.32 ± 6.55 | 37.43 [30.65, 39.50] | 0.852 ± 0.162 | 0.0034 ± 0.0033 |
| Faithful model | GS | 10.05 ± 3.54 | 9.68 [9.15, 11.35] | 0.144 ± 0.073 | 0.6736 ± 0.0482 |
| | GS (10,000 iterations) | 10.07 ± 3.57 | 9.76 [9.06, 11.30] | 0.146 ± 0.073 | 0.6700 ± 0.0444 |
| | GS (intensity target) | 9.63 ± 3.41 | 9.38 [8.38, 10.77] | 0.073 ± 0.025 | 0.7410 ± 0.0323 |
| | GD | 14.52 ± 3.36 | 14.80 [13.64, 16.29] | 0.257 ± 0.115 | 0.2479 ± 0.0734 |
| | GD (10,000 iterations) | 15.15 ± 3.42 | 15.45 [14.28, 16.86] | 0.281 ± 0.123 | 0.2175 ± 0.0793 |
| | GD (intensity target) | 14.49 ± 3.35 | 14.86 [13.60, 16.20] | 0.257 ± 0.115 | 0.2497 ± 0.0751 |
| | Transformer | 14.88 ± 3.43 | 14.98 [14.28, 16.66] | 0.276 ± 0.124 | 0.2309 ± 0.0803 |
| | Batched Transformer | 10.28 ± 3.38 | 10.08 [9.03, 11.78] | 0.065 ± 0.024 | 0.6393 ± 0.0454 |

**(b) Bench, single capture session.**

| **Model** | **Method** | **PSNR (dB) mean ± SD** | **PSNR median [IQR]** | **SSIM mean ± SD** | **NMSE mean ± SD** |
|---|---|---|---|---|---|
| Ideal model | GS | 15.88 ± 4.12 | 15.88 [14.05, 17.21] | 0.275 ± 0.146 | 0.1889 ± 0.0958 |
| | GS (10,000 iterations) | 15.91 ± 4.14 | 15.95 [14.05, 17.29] | 0.277 ± 0.147 | 0.1878 ± 0.0958 |
| | GD | 15.90 ± 4.26 | 16.07 [13.94, 17.24] | 0.270 ± 0.143 | 0.1870 ± 0.0921 |
| | GD (10,000 iterations) | 15.95 ± 4.23 | 16.10 [14.07, 17.33] | 0.276 ± 0.145 | 0.1854 ± 0.0929 |
| | GS+CITL (random init) | 14.90 ± 4.20 | 14.86 [13.01, 15.90] | 0.214 ± 0.137 | 0.2316 ± 0.1027 |
| | GS+CITL (sim-seeded) | 15.89 ± 4.14 | 15.93 [14.03, 17.31] | 0.276 ± 0.147 | 0.1879 ± 0.0925 |
| | GD+CITL (random init) | 18.02 ± 4.93 | 17.98 [15.77, 21.00] | 0.330 ± 0.136 | 0.1203 ± 0.0591 |
| | GD+CITL (sim-seeded) | 18.19 ± 4.86 | 18.48 [15.71, 21.01] | 0.328 ± 0.137 | 0.1168 ± 0.0599 |
| | Transformer | 15.95 ± 4.21 | 16.07 [14.03, 17.35] | 0.276 ± 0.145 | 0.1856 ± 0.0937 |
| | Batched Transformer | 15.72 ± 4.37 | 15.84 [13.74, 17.04] | 0.259 ± 0.144 | 0.1953 ± 0.0961 |
| | SAIL | 18.97 ± 5.02 | 19.53 [16.16, 22.17] | 0.376 ± 0.149 | 0.1028 ± 0.0613 |
| | Batched SAIL (750) | 18.64 ± 5.03 | 19.17 [15.85, 21.76] | 0.352 ± 0.152 | 0.1092 ± 0.0629 |
| | Batched SAIL (2000) | 19.11 ± 5.08 | 19.70 [16.19, 22.45] | 0.388 ± 0.153 | 0.1009 ± 0.0622 |
| Faithful model | GS | 15.44 ± 2.86 | 15.77 [13.12, 17.19] | 0.305 ± 0.119 | 0.2129 ± 0.1035 |
| | GS (10,000 iterations) | 15.45 ± 2.84 | 15.77 [13.14, 17.15] | 0.306 ± 0.118 | 0.2130 ± 0.1042 |
| | GD | 16.61 ± 2.91 | 16.73 [15.47, 18.12] | 0.268 ± 0.126 | 0.1606 ± 0.0790 |
| | GD (10,000 iterations) | 16.69 ± 2.91 | 16.84 [15.61, 18.23] | 0.274 ± 0.122 | 0.1588 ± 0.0816 |
| | GS+CITL (random init) | 14.75 ± 2.86 | 14.87 [12.62, 16.17] | 0.242 ± 0.116 | 0.2438 ± 0.0992 |
| | GS+CITL (sim-seeded) | 15.44 ± 2.85 | 15.76 [13.12, 17.18] | 0.305 ± 0.119 | 0.2132 ± 0.1039 |
| | GD+CITL (random init) | 18.89 ± 3.23 | 20.04 [18.14, 21.10] | 0.358 ± 0.125 | 0.1058 ± 0.0921 |
| | GD+CITL (sim-seeded) | 18.94 ± 3.23 | 20.04 [18.19, 21.17] | 0.357 ± 0.122 | 0.1051 ± 0.0926 |
| | Transformer | 16.68 ± 2.91 | 16.79 [15.58, 18.20] | 0.272 ± 0.123 | 0.1587 ± 0.0808 |
| | Batched Transformer | 15.55 ± 3.03 | 15.75 [13.81, 17.83] | 0.358 ± 0.118 | 0.2031 ± 0.0926 |
| | SAIL | 19.49 ± 3.50 | 20.75 [18.84, 22.03] | 0.383 ± 0.128 | 0.0977 ± 0.0963 |
| | Batched SAIL (750) | 19.16 ± 3.51 | 20.52 [18.58, 21.37] | 0.357 ± 0.131 | 0.1030 ± 0.0962 |
| | Batched SAIL (2000) | 19.51 ± 3.54 | 20.86 [18.91, 21.85] | 0.386 ± 0.129 | 0.0978 ± 0.0982 |

**Table 2 | Compute and amortization.** Cost of the 18-target campaign and of serving one further target, under the ideal (a) and faithful (b) forward models. Batched training is not cheaper than per-target training, since both spend one camera exposure per target per epoch. It yields one network instead of eighteen, which is what collapses the marginal cost of a new target. GS and GD have no weights to share, so the shared-network rows have no classical counterpart.

**(a) Ideal forward model.**

| Method | Networks produced | Campaign (hr) | Per target served |
|---|---|---|---|
| GD+CITL (random init) | none (no network) | 6.29 | 21.3 min |
| GD+CITL (sim-seeded) | none (no network) | 6.37 | 21.6 min |
| SAIL (per target) | 18 networks | 6.82 | 22.9 min |
| Batched SAIL (750 epochs) | 1 shared network | 8.35 | 0.015 s |
| Batched SAIL (2000 epochs) | 1 shared network | 23.03 | 0.015 s |

**(b) Faithful forward model.**

| Method | Networks produced | Campaign (hr) | Per target served |
|---|---|---|---|
| GD+CITL (random init) | none (no network) | 7.36 | 24.6 min |
| GD+CITL (sim-seeded) | none (no network) | 7.24 | 24.6 min |
| SAIL (per target) | 18 networks | 7.44 | 24.9 min |
| Batched SAIL (750 epochs) | 1 shared network | 8.99 | 0.015 s |
| Batched SAIL (2000 epochs) | 1 shared network | 25.16 | 0.015 s |

**Table 3 | Implementation settings.** All values are read from the archived run configurations of the reported experiments. (a) Learned networks. (b) Classical baselines, per method. (c) Hardware and scoring. (d) FNO baseline. (e) Training-regime map of what is inside each training loop and where each method is reported. Budget is epochs for the learned regimes and iterations for the iterative baselines.

**(a) Learned networks.** Weight decay and the learning-rate schedule are literals in the training scripts rather than configuration fields, and are reported from the script with the source recorded in the released analysis code. Every other value is read from the archived configuration snapshot.

| Setting | Transformer, per-target (sim.) | Transformer, batched (sim.) | SAIL (per-target) | Batched SAIL (750) | Batched SAIL (2000) |
|---|---|---|---|---|---|
| Epochs | 10000 | 10000 | 750 | 750 | 2000 |
| Batch size | 1 | 18 | 1 | - | - |
| Learning rate | 0.0001 | 0.0001 | 0.0001 | 0.0001 | 0.0001 |
| Patch size p | 500 | 500 | 500 | 500 | 500 |
| Heads | 16 | 16 | 16 | 16 | 16 |
| Encoder layers | 4 | 4 | 4 | 4 | 4 |
| d_model | 256 | 256 | 256 | 256 | 256 |
| d_ff | 1024 | 1024 | 1024 | 1024 | 1024 |
| Dropout | 0 | 0 | 0 | 0 | 0 |
| Pre-norm | yes | yes | yes | yes | yes |
| λ_sim | - | - | 0.1 | 0.1 | 0.1 |
| Optimiser | AdamW | AdamW | AdamW | AdamW | AdamW |
| Betas | (0.9, 0.999) | (0.9, 0.999) | (0.9, 0.999) | (0.9, 0.999) | (0.9, 0.999) |
| Weight decay | 1e-4 | 1e-4 | 1e-4 | 1e-4 | 1e-4 |
| LR schedule | ×0.95 every 1000 epochs | ×0.95 every 1000 epochs | ×0.95 every 250 epochs | ×0.95 every 250 epochs | ×0.95 every 250 epochs |
| Grid | 1000×1000 | 1000×1000 | 1000×1000 | 1000×1000 | 1000×1000 |
| Camera feedback | no | no | every epoch | every epoch | every epoch |

**(b) Classical baselines, per method.** Iterations, optimiser and learning rate are read from the archived run configurations, and the 10,000-iteration arm is the converged point of the same sweep. Only the CITL methods have two initializations.

| Method | Iterations | Optimiser | LR | Initialization | Camera |
|---|---|---|---|---|---|
| GS | 750 and 10,000 | alternating projection | -- | random | none |
| GD | 750 and 10,000 | adam | 0.03 | random | none |
| GS+CITL | 25 | alternating projection | -- | random, simulation-seeded | every iteration |
| GD+CITL | 750 | adam | 0.03 | random, simulation-seeded | every iteration |
| GD TV weight | 0 | | | | |

| | |
|---|---|
| Target formulation | amplitude |
| Grid | 1000 × 1000 |

**(c) Hardware and scoring.** The modulator, the fixed zero-order mask, the settle and calibration settings, and the two forward models, all shared by every bench/CITL/SAIL experiment. One manual rig calibration serves every bench capture set, so no method is scored through a different alignment or a different region of interest. The zero-order mask is applied identically to every method, so the excluded region is the same for all of them. Ideal and faithful are the two forward models of Methods 3, given here as pad factor, sinc envelope and fill factor.

| Setting | Value |
|---|---|
| SLM | 1080×1920 |
| DC block centre (px) | (495, 510) |
| DC block radius (px) | 20 |
| SLM settle time (s) | 0.3 |
| Calibration | manual, rotation -0.06° |
| Calibration burn-in (iter.) | 200 |
| Ideal forward model | pad 1, sinc off, fill 1 |
| Faithful forward model | pad 2, sinc on, fill 0.91 |

**(d) FNO baseline.** The standard unmodified FNO, run under the ideal forward model only. Mode coverage is full, so the operator is not truncated in frequency and the baseline is not handicapped by a spectral cut. Two conditions are reported: fno_scratch optimizes the phase against the target directly, and fno_regress fits the converged gradient-descent phase for the same target, which separates capacity from trainability. The learning rate was probed over five orders of magnitude and quality was insensitive to it, so the value quoted is representative.

| Setting | Value |
|---|---|
| Fourier modes | 500 × 500 (full coverage) |
| Width | 8 |
| Layers | 4 |
| Projection width | 128 |
| Activation | gelu |
| Domain padding | 0 |
| Epochs, from scratch | 10000 |
| Epochs, regression | 1500 |
| Learning rate | 0.01 (probed over 5 orders, quality insensitive) |
| Weight decay | 0.0001 |
| LR schedule | ×0.95 every 1000 epochs |
| Grid | 1000×1000 |
| Physics | ideal only (stated scope choice) |
| Conditions | fno_scratch, fno_regress (standard FNO only) |

**(e) Training-regime map.** Every method in the paper, classified by what sits inside its optimization loop, the budget it is given, how many targets one instance serves, how many instances are produced, and where it is reported. Learned regimes differ by whether the loop closes through the physical optics or through a simulated forward model. The classical rows carry iteration counts rather than epochs because they train nothing and produce no network. Counts under Networks produced are per forward model, so a per-target regime yields 18 and a batched regime yields one. The final column is the audit trail: every figure and table in which each regime appears.

| Regime | In the training loop | Budget | Targets per network | Networks produced | Reported in |
|---|---|---|---|---|---|
| Transformer, per-target (simulation) | simulated forward model | 10000 | 1 | 18 per physics | Fig 2, Fig 3, Fig 4, E3a, E3c, E5, E8, Table 1 |
| Transformer, batched (simulation) | simulated forward model | 10000 | 18 | 1 per physics | Fig 2, Fig 3, E3a, E3c, Table 1 |
| SAIL (per-target) | physical optics, every epoch | 750 | 1 | 18 per physics | Fig 3, Fig 4, Fig 5, E3a, E3c, E8, Table 1, Table 2 |
| Batched SAIL, 750 epochs | physical optics, every epoch | 750 | 18 | 1 per physics | Fig 5, Table 1, Table 2 |
| Batched SAIL, 2000 epochs | physical optics, every epoch | 2000 | 18 | 1 per physics | Fig 4, Fig 5, E3a, E3c, E8, Table 1, Table 2 |
| GS | none (no training) | 750 iterations | -- | none | Fig 2, Fig 3, Fig 4, E4, E5, E8, Table 1 |
| GS, converged | none (no training) | 10,000 iterations | -- | none | E5, Table 1 |
| GD | none (no training) | 750 iterations | -- | none | Fig 2, Fig 3, Fig 4, E4, E5, E8, Table 1 |
| GD, converged | none (no training) | 10,000 iterations | -- | none | E5, Table 1 |
| GS+CITL, random init | physical optics, every iteration | 25 iterations | -- | none | Fig 3, Table 1 |
| GS+CITL, simulation-seeded | physical optics, every iteration | 25 iterations | -- | none | Fig 3, Table 1 |
| GD+CITL, random init | physical optics, every iteration | 750 iterations | -- | none | Fig 3, Table 1, Table 2 |
| GD+CITL, simulation-seeded | physical optics, every iteration | 750 iterations | -- | none | Fig 3, Fig 4, E8, Table 1, Table 2 |

## Extended data legends

**Extended Data Fig. E1 | FNO baseline.** A standard, unmodified FNO trained per target under the ideal model. ***a***, Reconstruction quality of the two training arms, against GD and the transformer. Trained from scratch the operator reaches 8.75 dB median PSNR over the 18 targets. Trained by regression against converged GD phases it reaches 37.43 dB, matching converged GD quality, against 52.13 dB for the transformer. ***b***, Learning-rate probe for the from-scratch arm, which is insensitive to learning rate across five orders of magnitude, so the failure is not a tuning artifact. With 256.0M parameters against the transformer's 195.7M, the gap is one of trainability under the far-field objective, not capacity. The regression control converged within its 1,500-epoch budget, while the from-scratch arm received the full 10,000.

**Extended Data Fig. E2 | Patch-size sweep.** ***a***, Median PSNR under the ideal model falls monotonically on 18 of 18 targets as patch size shrinks, from 53.22 dB at p = 500 to 36.19 at p = 250, 23.75 at p = 125, and 11.80 at p = 50. ***b***, The mechanism. The output head is a single linear map from the 256-dimensional token code to $2p^2$ pixel values, so it shrinks a hundredfold across the sweep, from 128 million parameters at p = 500 to 1.28 million at p = 50, while attention only sees more tokens. The head, not the token count, is the binding constraint, and attention over fewer, larger, fully global tokens is therefore the operative design point in the far field. Patch size is varied with width, depth and training schedule fixed, and under the ideal forward model only. Degraded performance in simulation at small p does not establish that the same range holds on the bench, where camera feedback may make finer patches useful. That, and a joint sweep over patch size and width, are future work.

**Extended Data Fig. E3 | Attention maps and bias-dominance control.** ***a***, Mean attention maps, one per training regime. At p = 500 the 1000 × 1000 field is four tokens, so each map is the 4 × 4 token-to-token attention matrix, averaged over every head, every layer and all 18 targets, on a shared scale with the uniform value 0.25 marked. Per-target training sits near uniform, and requiring one network to serve all 18 commits the heads. The corresponding selectivity values are in Fig. 5c, and the batched regimes shown here are the 2,000-epoch budget. ***b***, The bias-dominance control. Evaluating the trained network with its input-dependent pathway removed, so that only the learned bias reaches the output, collapses reconstruction from 52.13 dB median (per-target, simulation) to 8.00 dB, a paired fall of 43.83 dB on 18 of 18 targets, and from 27.66 dB to 8.00 dB batched, a fall of 19.90 dB, also on 18 of 18. The network's quality lives in its input-dependent computation, not in a memorized offset. ***c***, The same attention maps target by target, one row per training regime and one map per target, on the shared scale of panel ***a***. The averaging in panel ***a*** hides nothing. Per-target training stays near the uniform value on every target, and the batched networks are selective on every target, so the training regime rather than the individual picture sets the behavior. These are simulation-only by design because the diagnostic evaluates checkpoints through the simulated forward model.

**Extended Data Fig. E4 | Target formulation, amplitude versus intensity.** Paired per-target differences, amplitude minus intensity, for GS and GD run under both target formulations on all 18 targets. ***a***, Ideal forward model. ***b***, Faithful forward model. For GD the comparison is an equivalence control. The implementation squares the amplitude back to intensity before the loss is formed, so both parameterizations optimize the same intensity-domain objective, and the paired differences are consistent with run-to-run variation (+0.19 dB median ideal, $P = 0.28$, and +0.05 dB faithful, $P = 0.20$, neither significant). For GS the formulation changes the algorithm because the chosen array is substituted as the replay-plane magnitude constraint. Raw intensity substitution distorts brightness quadratically, costing 21.44 dB median under the ideal model (worse on 17 of 18 targets, $P < 0.001$) and 0.45 dB under the faithful model (16 of 18, P <

0.001), with near-binary targets unaffected and continuous-tone scenes collapsing, the signature of the mechanism. The published amplitude formulation is therefore the stronger configuration for GS and an equivalent parameterization for GD.

**Extended Data Fig. E5 | Convergence and runtime of iterative methods compared to single-shot self-attention.** Median PSNR over the 18 targets for GS and GD under the ideal (a, b) and faithful (c, d) forward models, with shading spanning the interquartile range in all four panels. a, c, Quality against iteration count. The dotted line marks the 750-iteration operating point reported in the main text and the dashed line marks the per-target transformer, reached in one forward pass. Every point is an independent run from a random initialization, so step-to-step wobble is run-to-run scatter rather than a trajectory. GD overtakes the transformer near 5,000 iterations under the ideal model and finishes at 56.49 dB median against 52.13 dB at 10,000 iterations. Under the faithful model GD plateaus at 15.45 dB against the transformer's 14.98 dB and GS stays below 10 dB, so every simulation-only method lands together. b, d, The same quality against median wall-clock seconds per target. The interquartile band is drawn at the median-time abscissa because each target reaches a given budget at a slightly different time. Matching the transformer's ideal-model quality takes GD roughly 5,000 iterations and about 15 s per target, against 12 to 24 ms for one forward pass.

**Extended Data Fig. E6 | Generalization to unseen targets.** Each column is one target pattern that was not in the training set. Rows are, from top, the desired target far field, the far field reconstructed from the predicted hologram, and the predicted phase, which spans -π to π. The network is a compact 3.3-million-parameter proof-of-concept model at the native 28 × 28 grid, with a patch size of 14 so that the field is again four tokens, trained on a narrow curriculum and then asked for geometric shapes and free-form patterns it had never seen (Methods, Section 9). Panels are the study's archived outputs, shown display-scaled. No PSNR is printed on them because only canonically scored values are quoted in this work. The resolution is a computational choice, not a limit of generalizability.

**Extended Data Fig. E7 | Optical setup. A fiber-coupled**
650 nm laser is collimated (f = 50 mm) and linearly polarized before being directed to a reflective LCoS Holoeye LUNA SLM via a 50:50 beam splitter. The phase-modulated wavefront is reflected and relayed to a Fourier lens (f = 100 mm), forming the far-field intensity at the focal plane, which is recorded by a Canon digital camera. Red shading indicates the optical propagation path.

**Extended Data Fig. E8 | Qualitative reconstructions with regions of interest.** Full-frame reconstructions for the six methods of Fig. 4 from the single-session bench capture, shown on a shared display scale with canonical full-frame PSNR values. Enlarged regions of interest (ROI) highlight fine structure. Five targets are shown instead of all eighteen, chosen to span the range of SAIL's margin over its strongest adapted comparator, simulation-seeded GD+CITL, on the bench under the ideal forward model. That margin is positive on all 18 targets and runs from +0.10 dB to +1.58 dB, and the five are the smallest, the largest and three approximately evenly spaced between them. Each is shown under both forward models, as a full-frame grid and as regions of interest, at the full resolution of the analysis output, which is what makes showing the range at that resolution practical. The remaining thirteen targets are in the deposit (Supplementary Information). ROI coordinates are versioned with the released code. The GD+CITL arm shown is the simulation-seeded one and Batched SAIL is the 2,000-epoch budget.

**Extended Data Fig. E9 | Aberration-severity sweep.** Fourier lens displaced toward the camera by nominal caliper settings of 0.05, 0.1, 1 and 2 mm (uncertainty approximately ±0.05 mm), one target, one frozen calibration, fixed exposure. The three arms are the never-adapted

simulation hologram, SAIL trained on the aligned bench, and SAIL retrained at each displacement. PSNR spans only about 0.5 dB across the sweep on this target while SSIM moves by a factor of 2.3, so SSIM carries the sensitivity. Retrained SAIL does not recover the aligned baseline (-0.07 to -0.38 dB as displacement grows), and SAIL's SSIM advantage over the never-adapted arm decays from 0.059 to 0.002. Defocus removes quality that in-loop training cannot buy back. The converse control carries the specificity result. The hologram trained at 2 mm and displayed on the aligned system scores 0.61 dB below the aligned-trained one, roughly fifteen times the 0.04 dB measurement floor. Return-to-zero checks passed within +0.03 and +0.04 dB against a day-scale rig reproducibility of 0.14 dB.

**Extended Data Fig. E10 | Why the propagation regime changes the architecture.** Illustrative simulation on one canonical target, matched in every respect other than propagation. Both regimes use the same GS algorithm, the same 60-iteration budget and the same zero-phase initialization, on a 1000 × 1000 modulator at the rig's 4.5 μm pitch under 650 nm illumination. ***a***, Phase-only holograms of the same target computed under Fresnel propagation by the angular spectrum method at z = 10 cm, the order of the working distance of the near-eye systems cited in the Introduction, and under Fraunhofer propagation. The near-field phase visibly retains the spatial layout of the scene. The far-field phase is structureless at every scale. ***b***, The locality probe. A constant π offset is added to one 100 × 100 pixel patch of each hologram, 0.5 mm at the modulator, and the resulting change in reconstructed intensity is imaged. Under Fresnel propagation the disturbance stays compact and sits at the corresponding location in the image. Under Fraunhofer propagation it spreads across the whole reconstruction because a patch of a far-field hologram is not a place in the image but a band of plane-wave components of all of it. Within a box three patch widths across, the near-field disturbance carries 24.3% of the total change, against the 9.0% expected if it were spread uniformly, while the far-field disturbance carries 9.2% and is therefore indistinguishable from uniform spreading. A constant offset is used because a random patch radiates into the modulator's full angular bandwidth in any regime and would probe the pixel pitch rather than the propagation physics. This figure is an illustrative simulation and enters no quantitative claim in this work.

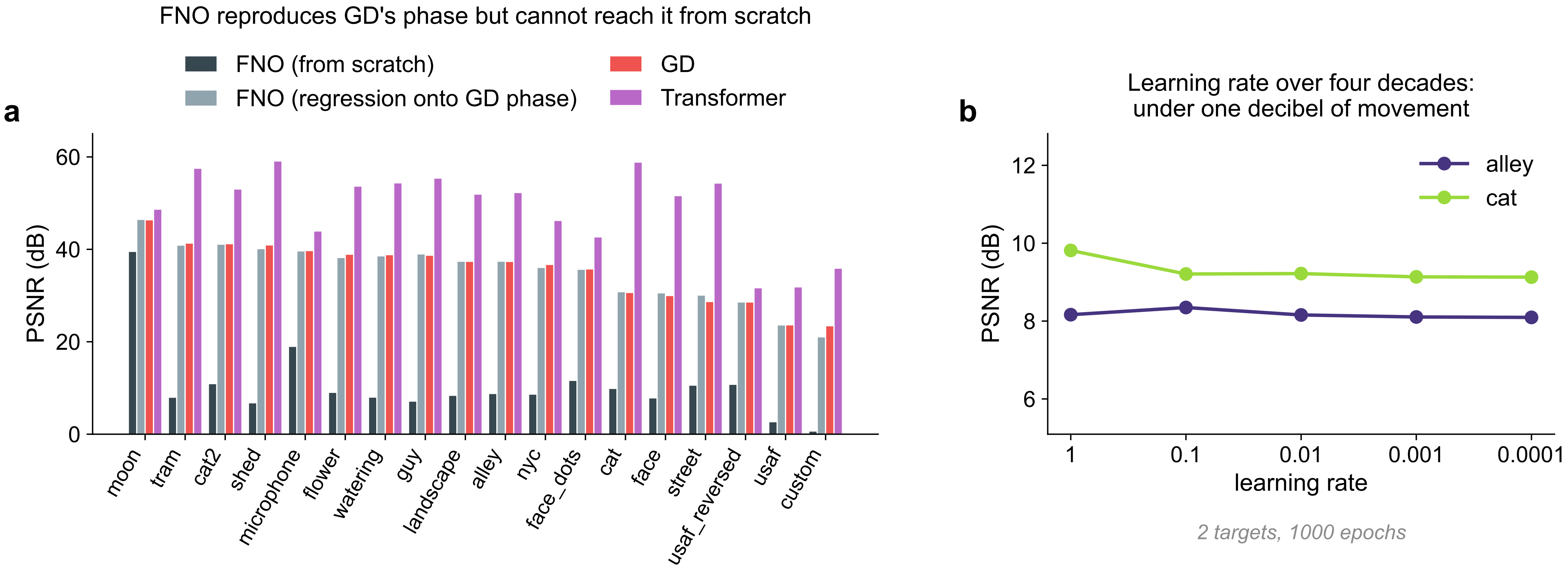

FNO reproduces GD's phase but cannot reach it from scratch
FNO (from scratch)
FNO (regression onto GD phase)
GD
Transformer
a
PSNR (dB)
0
20
40
60
moon
tram
cat2
shed
microphone
flower
watering
guy
landscape
alley
nyc
face_dots
cat
face
street
usaf_reversed
usaf
custom
b
Learning rate over four decades:
under one decibel of movement
alley
cat
PSNR (dB)
6
8
10
12
1
0.1
0.01
0.001
0.0001
learning rate
2 targets, 1000 epochs

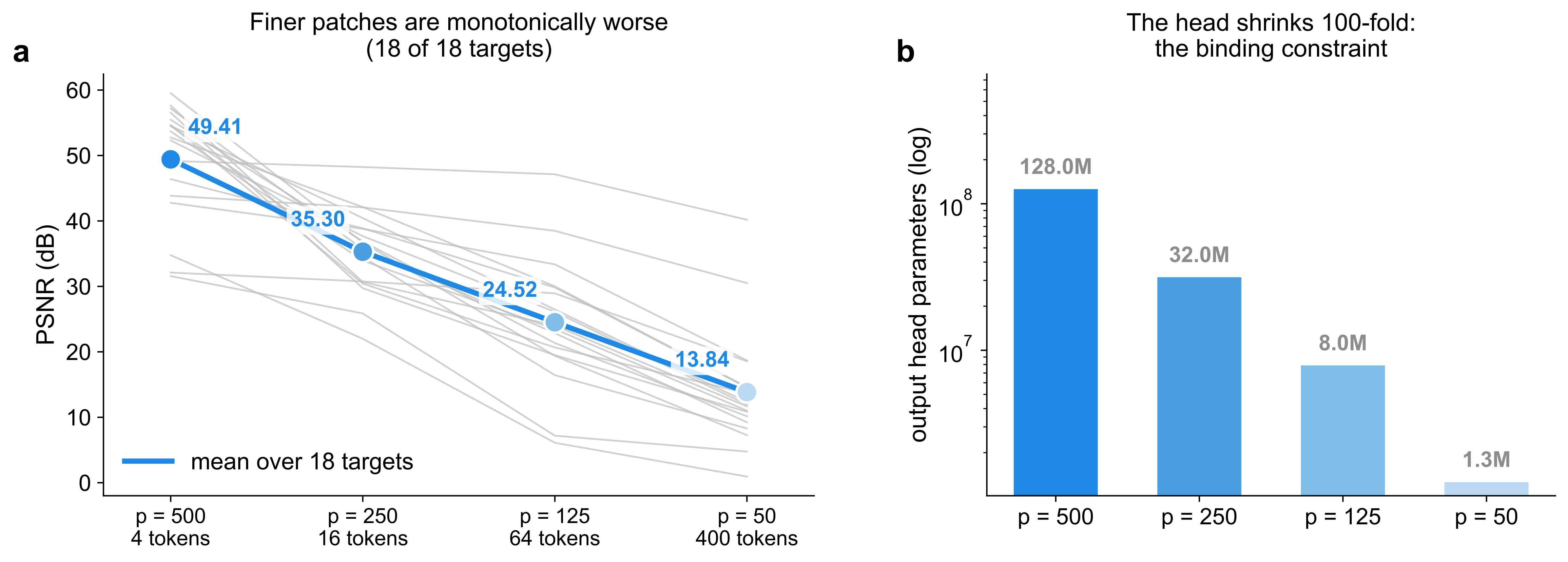
a
Finer patches are monotonically worse
(18 of 18 targets)
PSNR (dB)
60
50
40
30
20
10
0
49.41
35.30
24.52
13.84
mean over 18 targets
p = 500
4 tokens
p = 250
16 tokens
p = 125
64 tokens
p = 50
400 tokens
b
The head shrinks 100-fold:
the binding constraint
output head parameters (log)
$10^8$
$10^7$
128.0M
32.0M
8.0M
1.3M
p = 500
p = 250
p = 125
p = 50

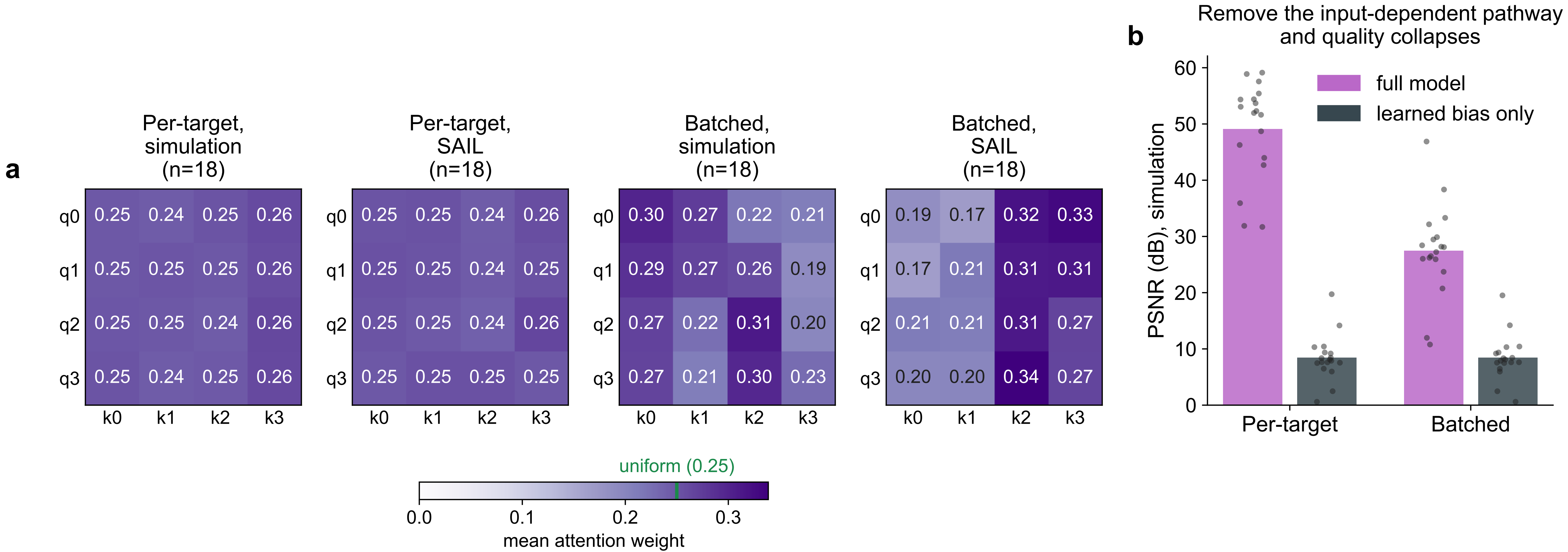
a
Per-target, simulation (n=18)
Per-target, SAIL (n=18)
Batched, simulation (n=18)
Batched, SAIL (n=18)
q0
q1
q2
q3
k0
k1
k2
k3
uniform (0.25)
0.0
0.1
0.2
0.3
mean attention weight
b
Remove the input-dependent pathway and quality collapses
full model
learned bias only
PSNR (dB), simulation
Per-target
Batched

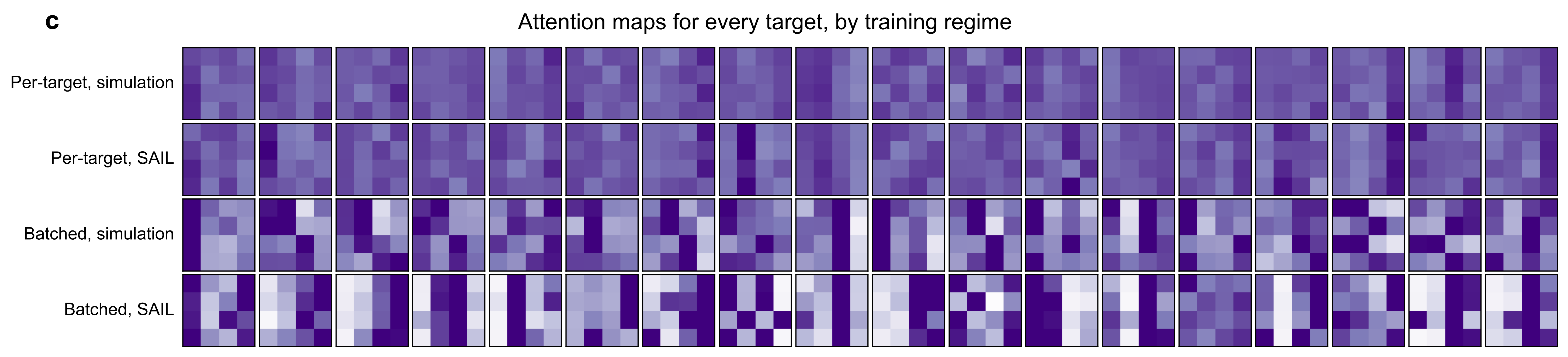
c
Attention maps for every target, by training regime
Per-target, simulation
Per-target, SAIL
Batched, simulation
Batched, SAIL

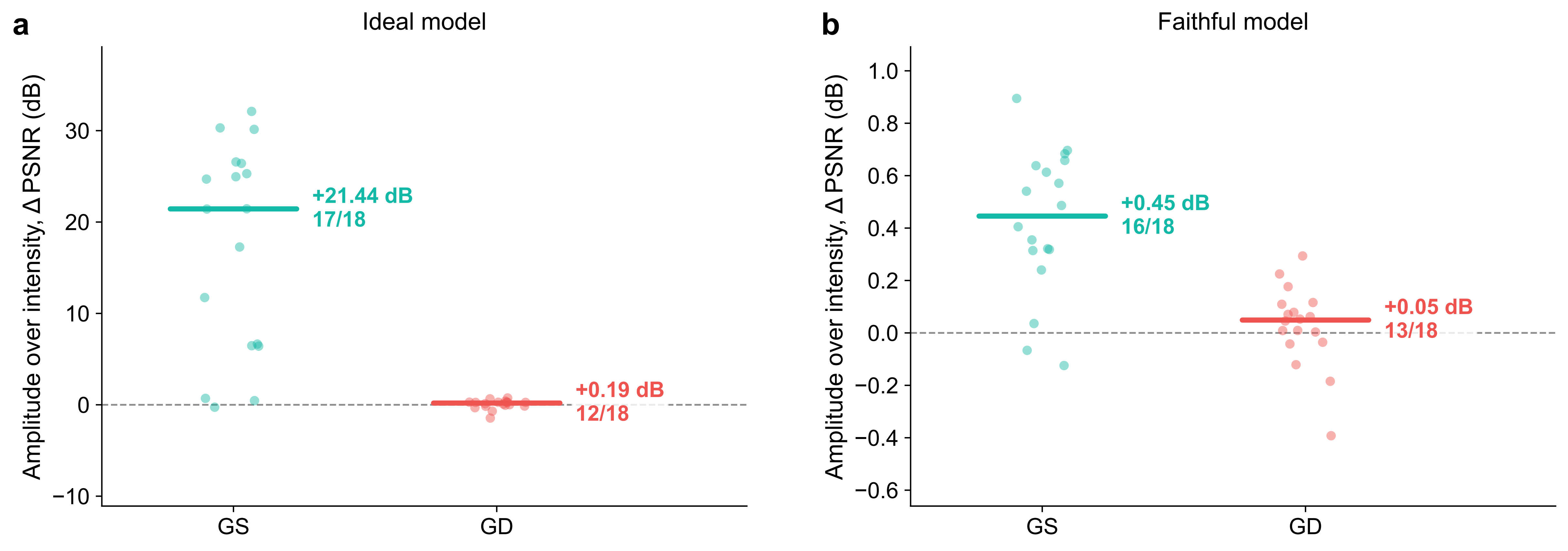
a
Ideal model
Amplitude over intensity, Δ PSNR (dB)
30
20
10
0
−10
+21.44 dB
17/18
+0.19 dB
12/18
GS
GD
b
Faithful model
Amplitude over intensity, Δ PSNR (dB)
1.0
0.8
0.6
0.4
0.2
0.0
−0.2
−0.4
−0.6
+0.45 dB
16/18
+0.05 dB
13/18
GS
GD

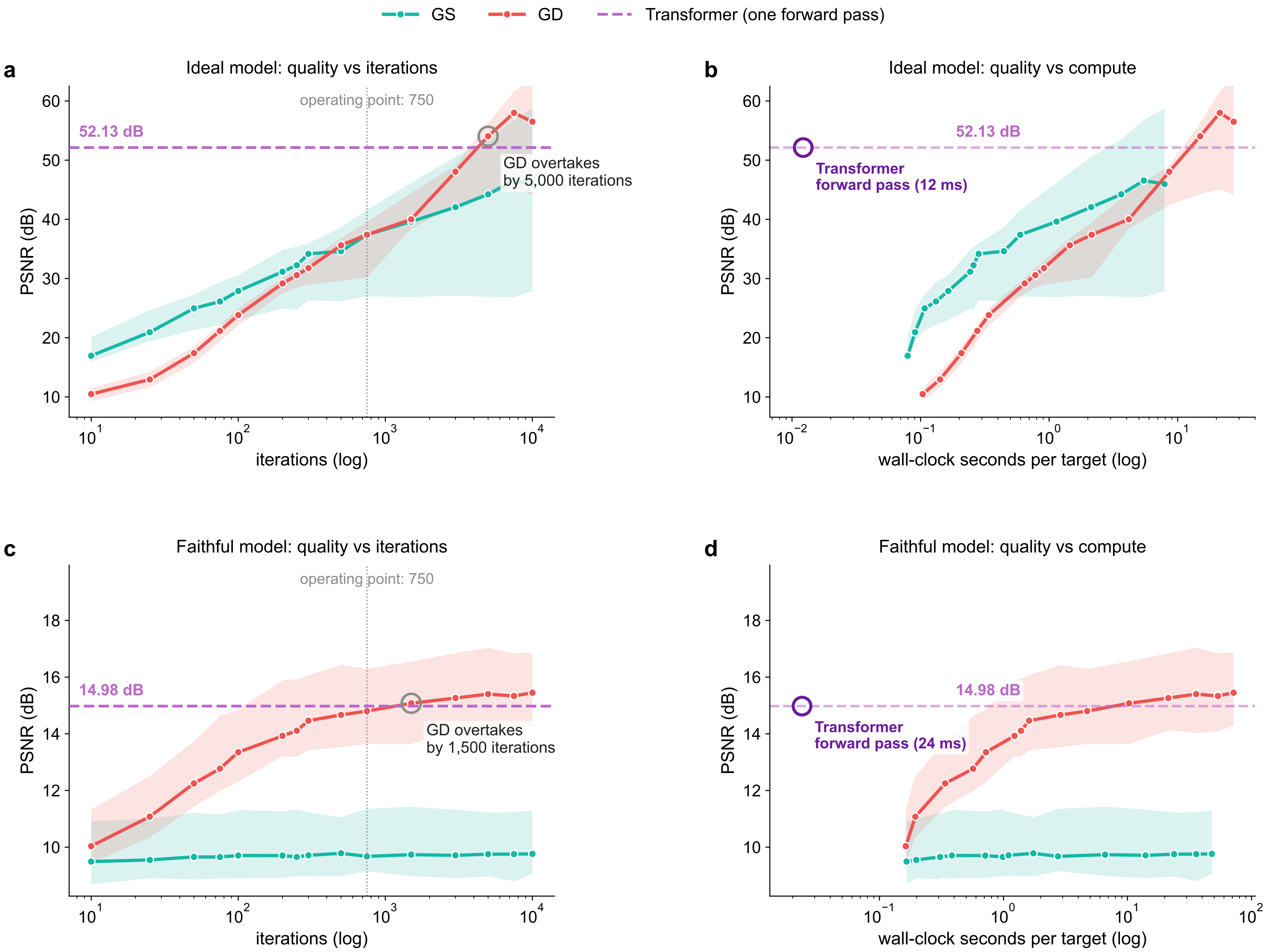

GS
GD
Transformer (one forward pass)
a
Ideal model: quality vs iterations
operating point: 750
52.13 dB
GD overtakes
by 5,000 iterations
PSNR (dB)
iterations (log)
b
Ideal model: quality vs compute
52.13 dB
Transformer
forward pass (12 ms)
PSNR (dB)
wall-clock seconds per target (log)
c
Faithful model: quality vs iterations
operating point: 750
14.98 dB
GD overtakes
by 1,500 iterations
PSNR (dB)
iterations (log)
d
Faithful model: quality vs compute
14.98 dB
Transformer
forward pass (24 ms)
PSNR (dB)
wall-clock seconds per target (log)

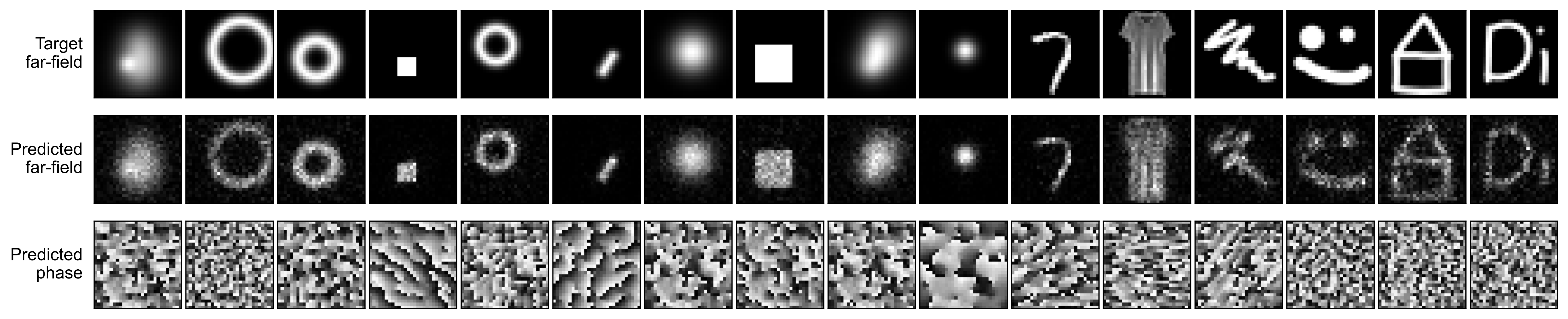
Target far-field
Predicted far-field
Predicted phase

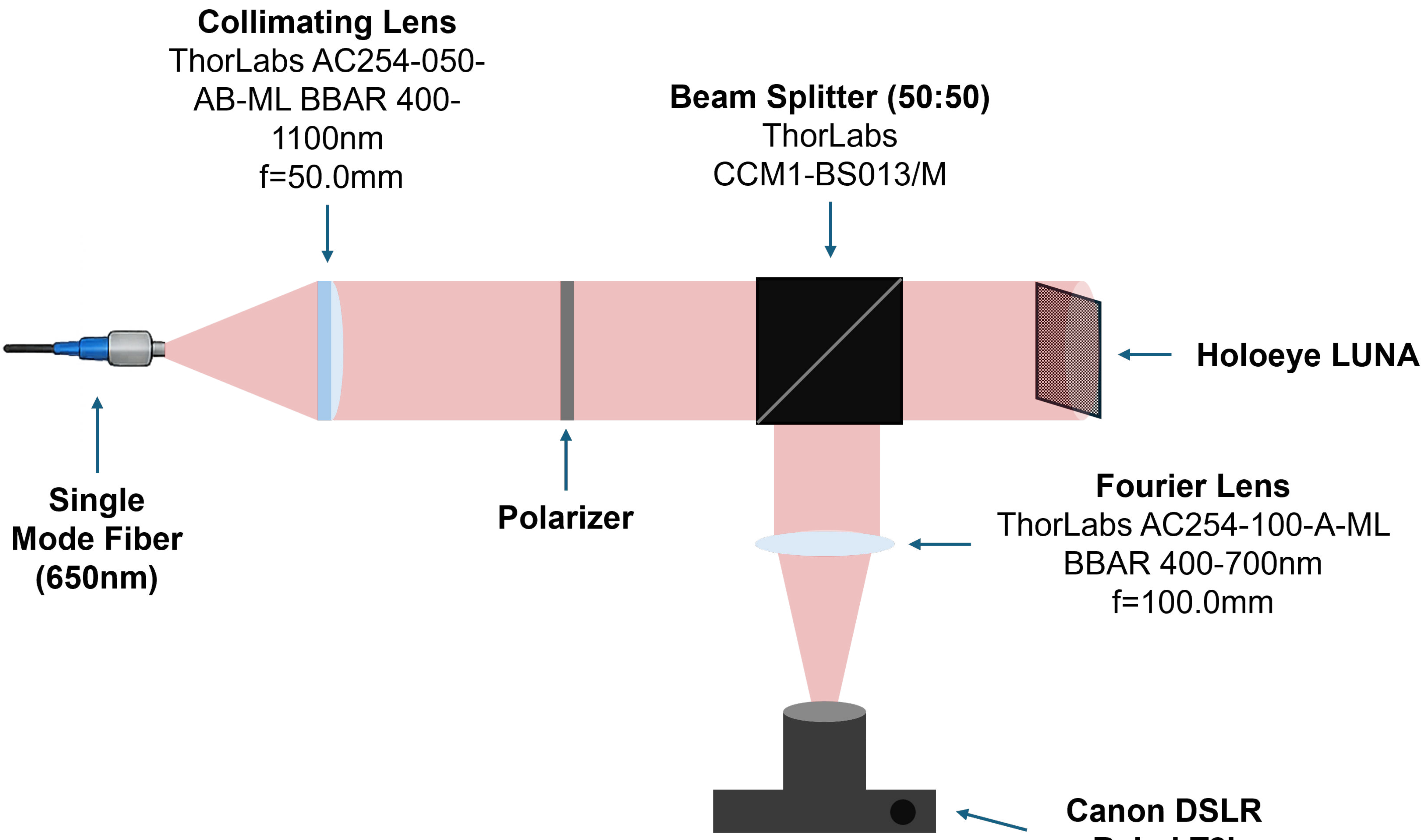
Collimating Lens
ThorLabs AC254-050-
AB-ML BBAR 400-
1100nm
f=50.0mm
Beam Splitter (50:50)
ThorLabs
CCM1-BS013/M
Holoeye LUNA
Single
Mode Fiber
(650nm)
Polarizer
Fourier Lens
ThorLabs AC254-100-A-ML
BBAR 400-700nm
f=100.0mm
Canon DSLR
Rebel T2i

# Supplementary Information section

## Extended Data Fig. E8 | Qualitative reconstructions with regions of interest

Full-frame reconstructions for every method from the single-session bench capture, shown on a shared display scale with canonical full-frame PSNR values. Enlarged regions of interest (ROI) highlight fine structure.

Five targets are shown rather than all eighteen, chosen to span the range of SAIL's margin over its main comparator, simulation-seeded GD+CITL. That margin is positive on all 18 targets and runs from +0.10 dB to +1.58 dB, and the five are the smallest, the largest and three approximately evenly spaced between them. The selection and the order below are set by that margin under the IDEAL forward model. Both forward models are then reported for every one of the five, as a full-frame grid and as regions of interest, at the full resolution the analysis wrote. The PSNR values printed on the panels are each condition's own canonical score, so the faithful-model panels carry faithful-model values and are not comparable with the ordering margin. Showing the range rather than every target is what makes full resolution practical here. The remaining thirteen targets are in the deposit. Every page is labelled with its target and forward model, so no page depends on its position in the sequence.

ROI coordinates are versioned with the released code (roi.json). The unmodulated zero-order component is excluded from scoring by a fixed mask, identical for every method, and retained for visualization (Methods).

In order of ideal-model margin: usaf reversed (+0.10 dB), guy (+0.51 dB), cat (+0.89 dB), watering (+1.23 dB), cat2 (+1.58 dB).

**usaf reversed, ideal**

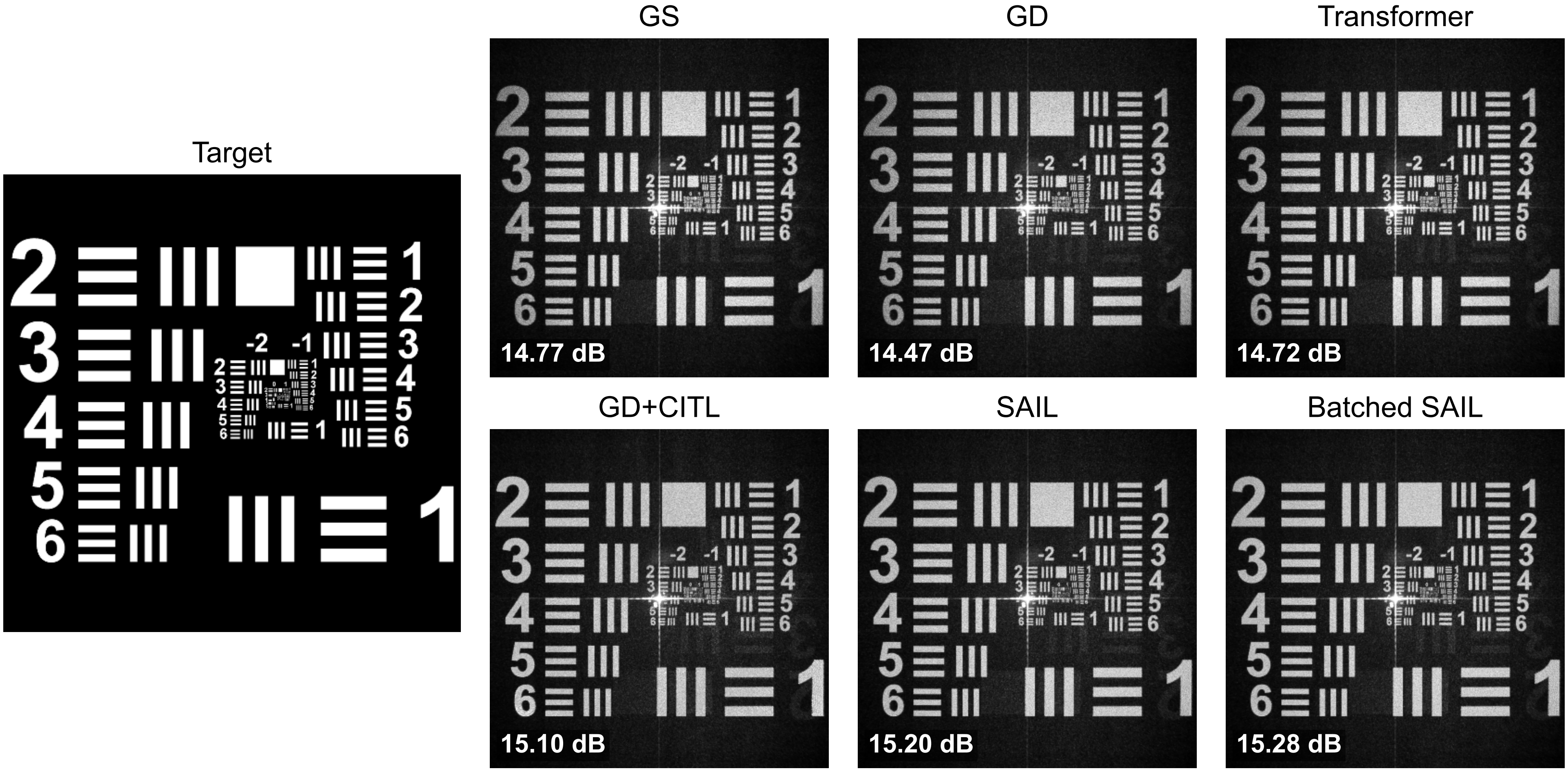

**usaf reversed, ideal, regions of interest**

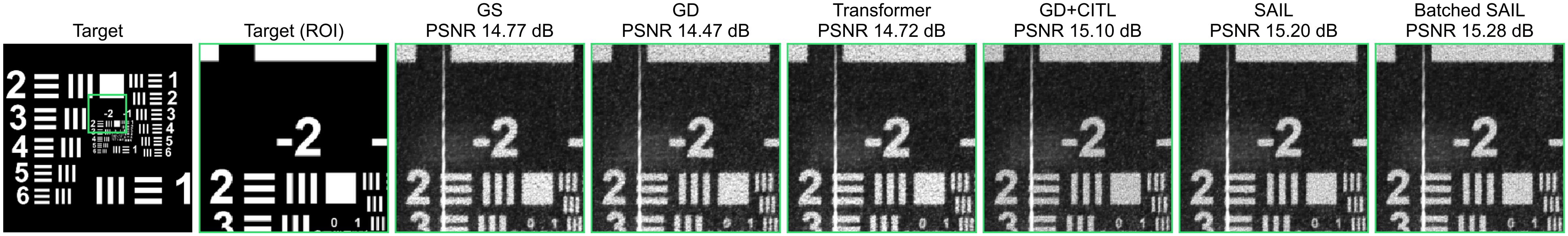

**usaf reversed, faithful**

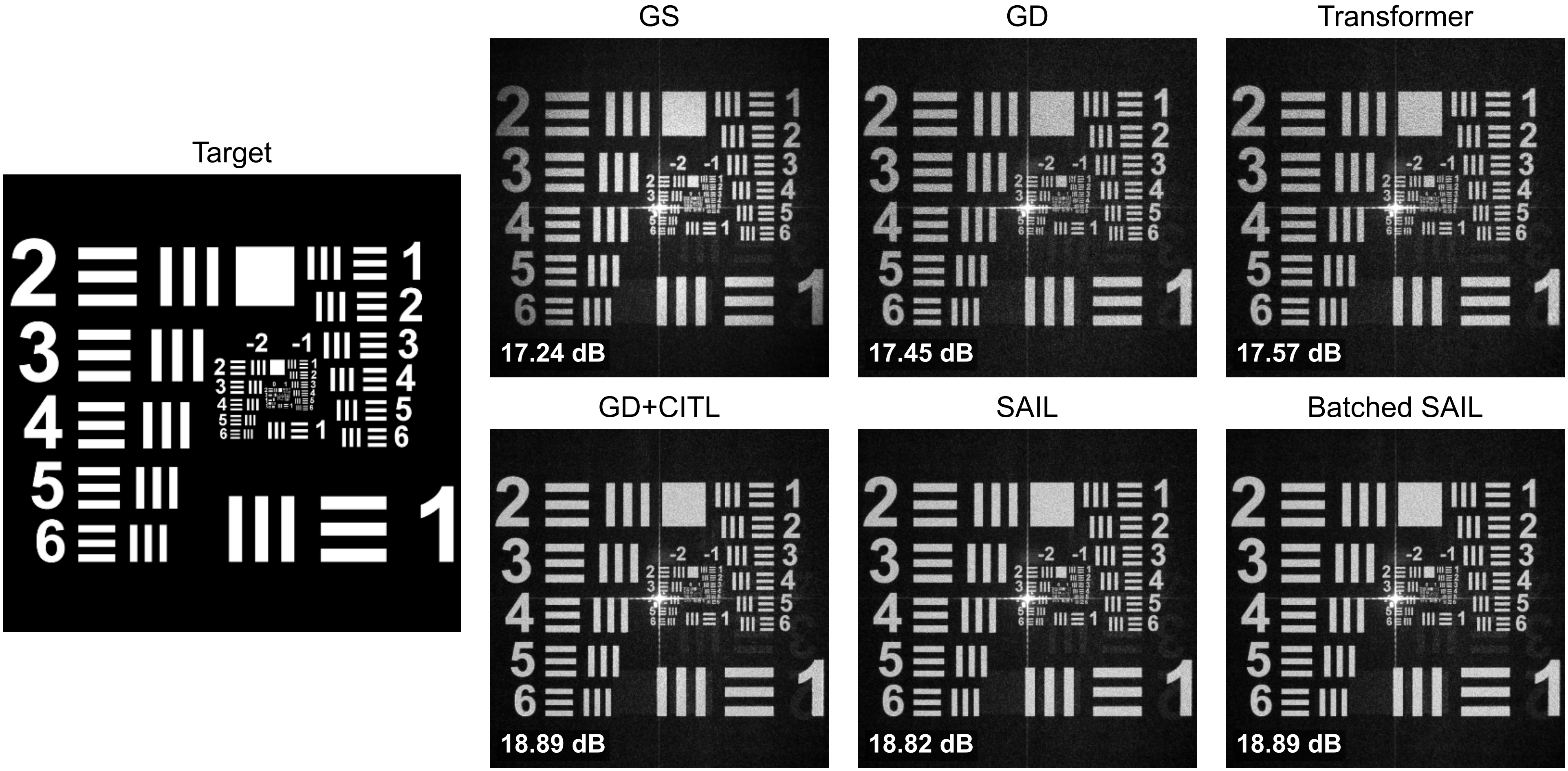

## usaf reversed, faithful, regions of interest

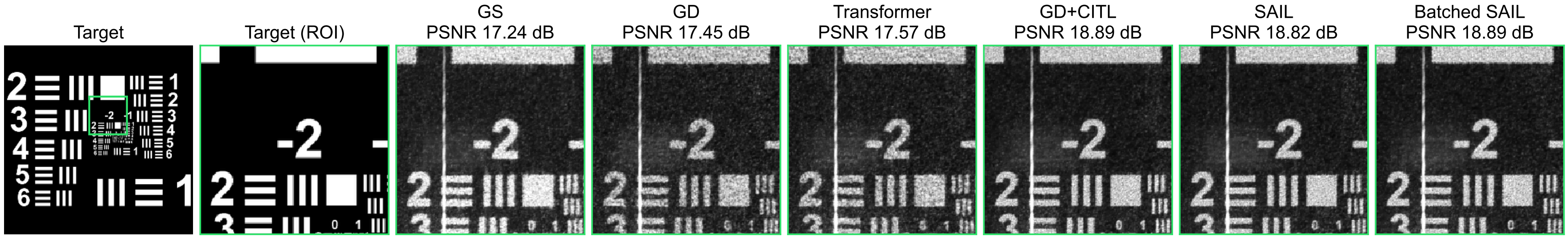

**guy, ideal**

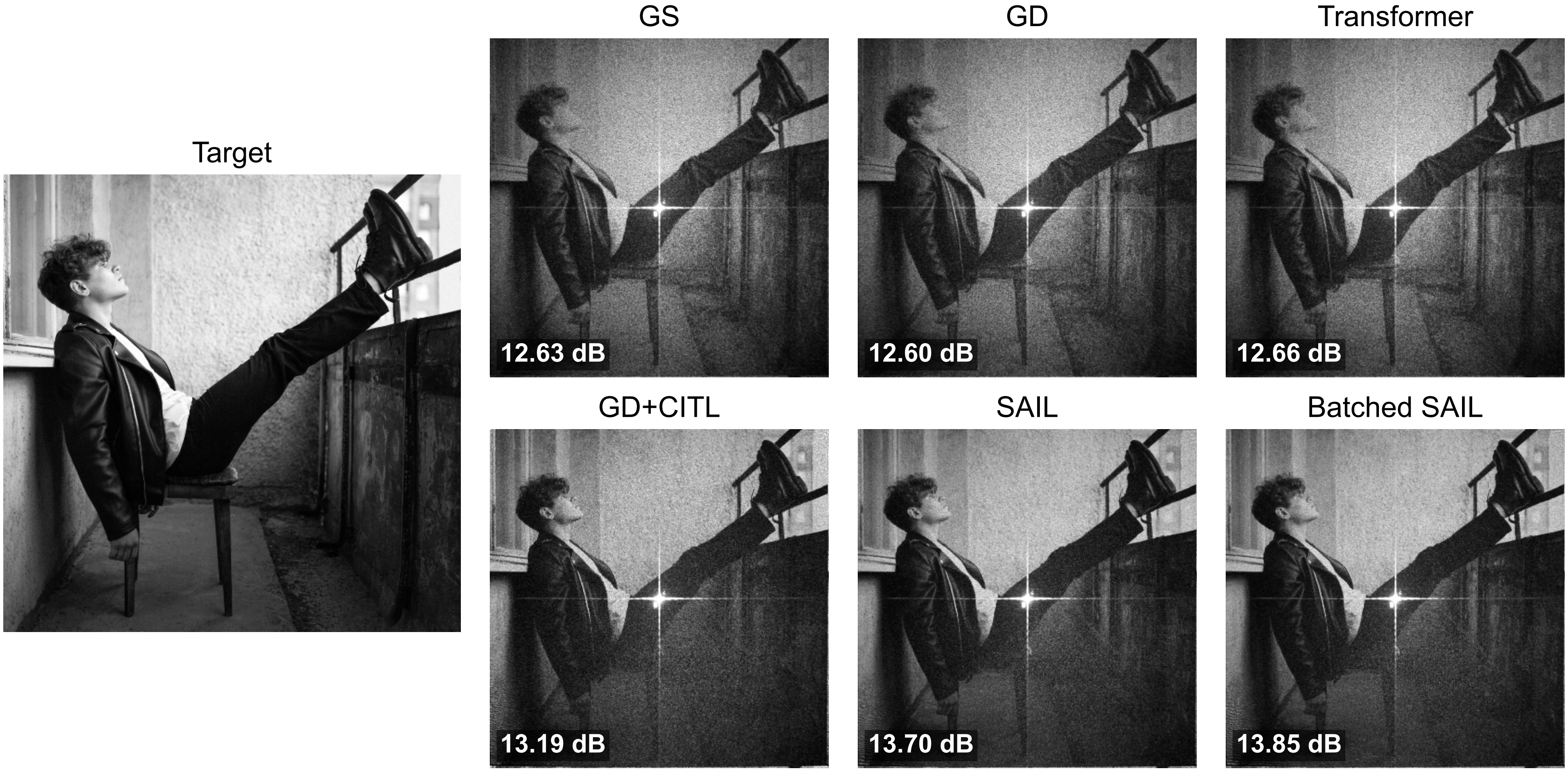
Target
GS
12.63 dB
GD
12.60 dB
Transformer
12.66 dB
GD+CITL
13.19 dB
SAIL
13.70 dB
Batched SAIL
13.85 dB

**guy, ideal, regions of interest**

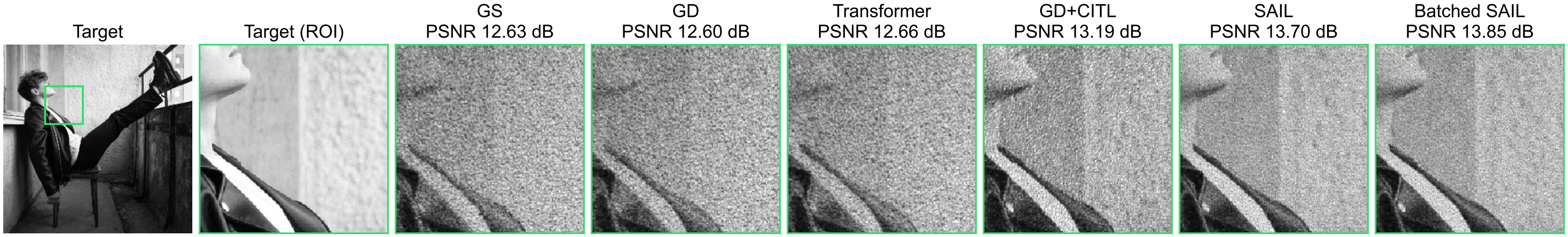

**guy, faithful**

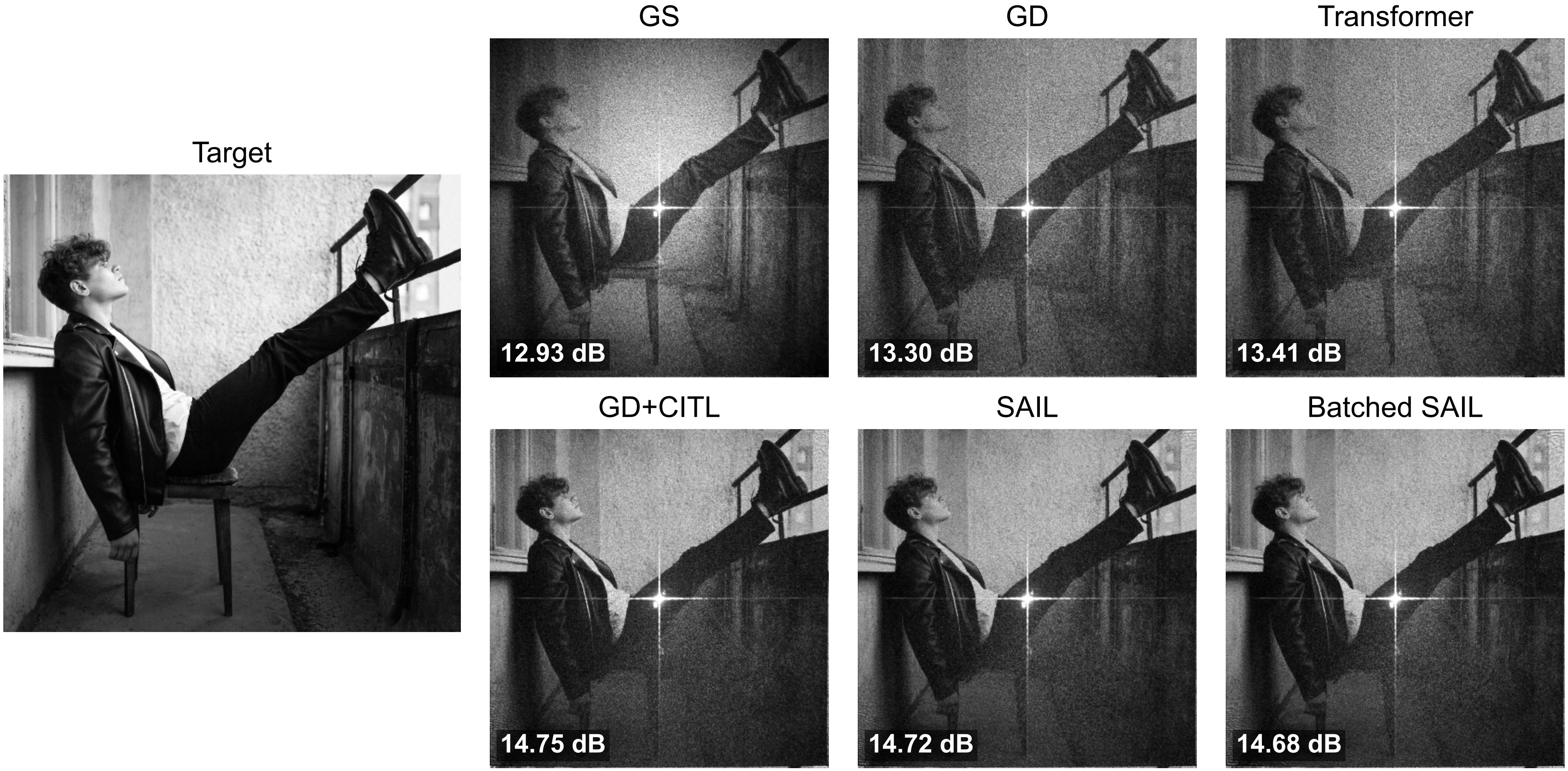

## guy, faithful, regions of interest

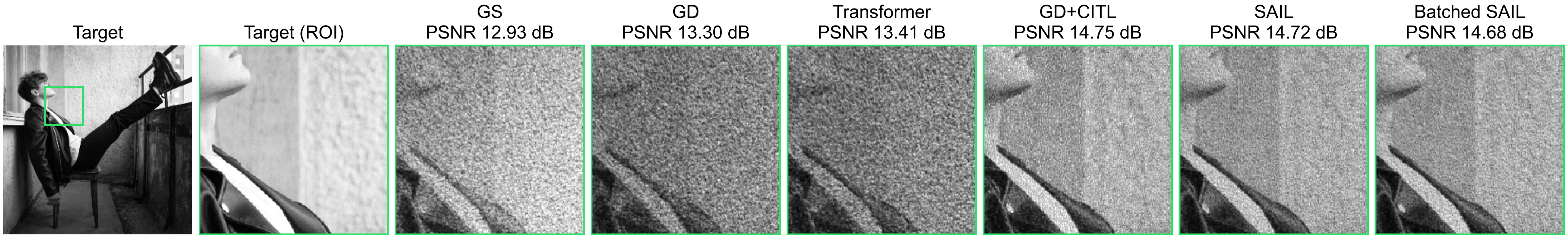

**cat, ideal**

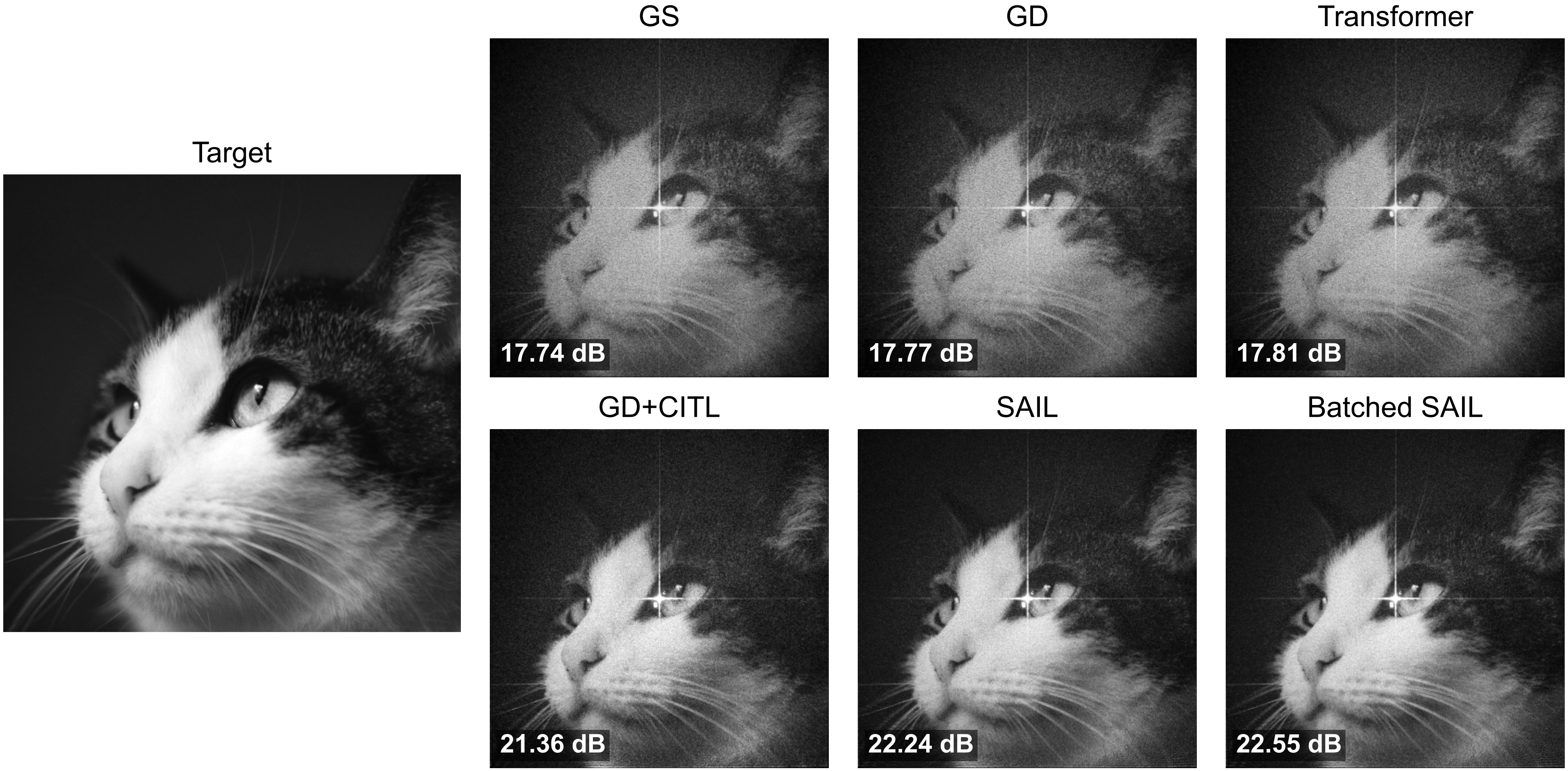

Target
GS
17.74 dB
GD
17.77 dB
Transformer
17.81 dB
GD+CITL
21.36 dB
SAIL
22.24 dB
Batched SAIL
22.55 dB

**cat, ideal, regions of interest**

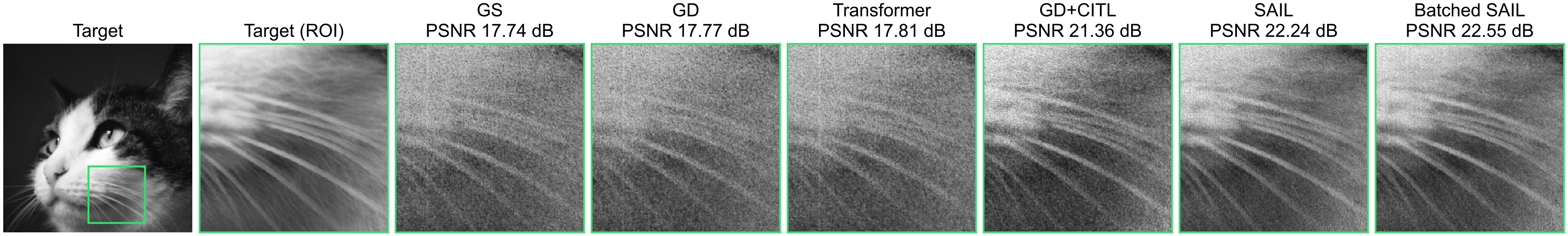

**cat, faithful**

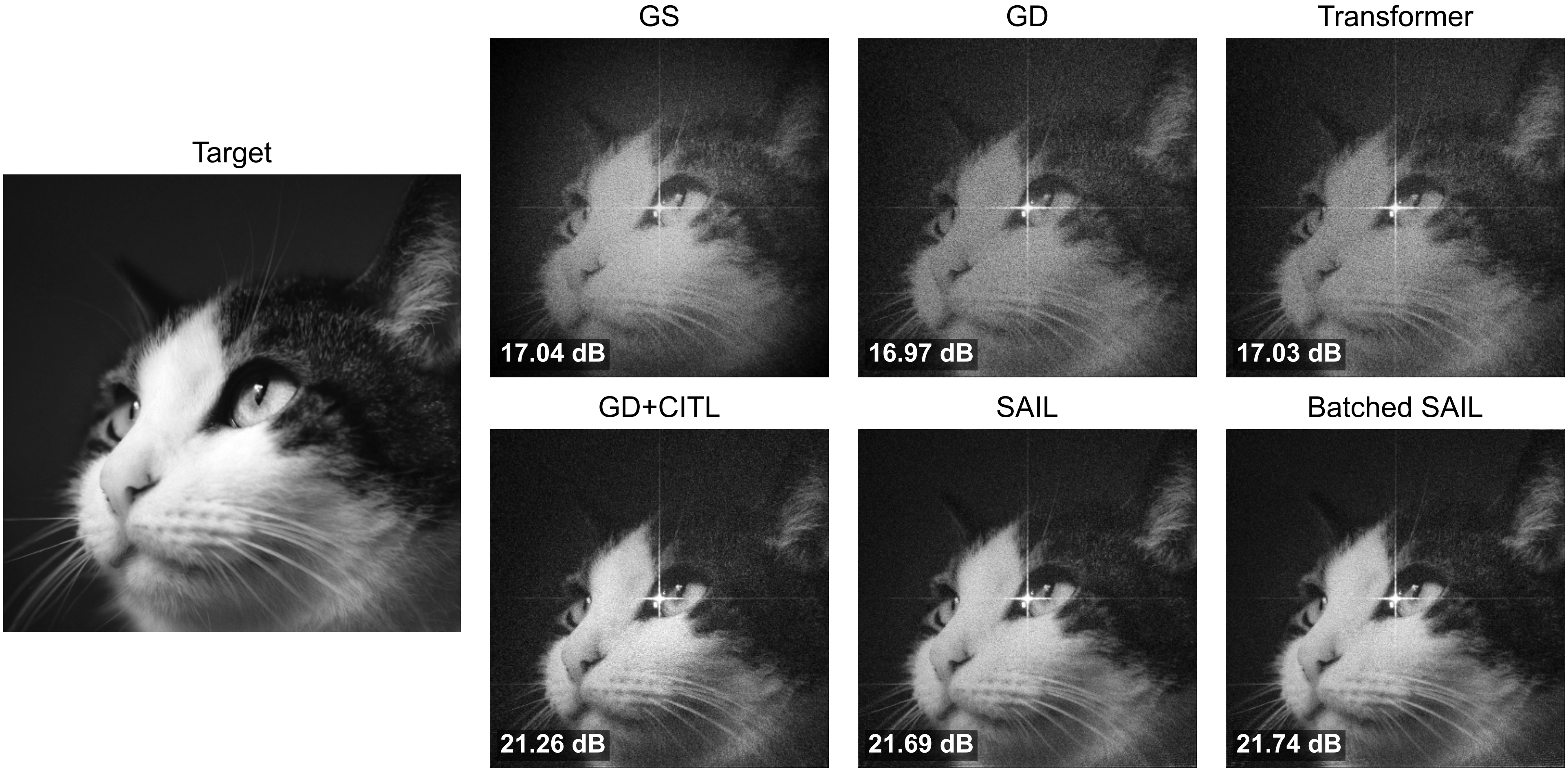

**cat, faithful, regions of interest**

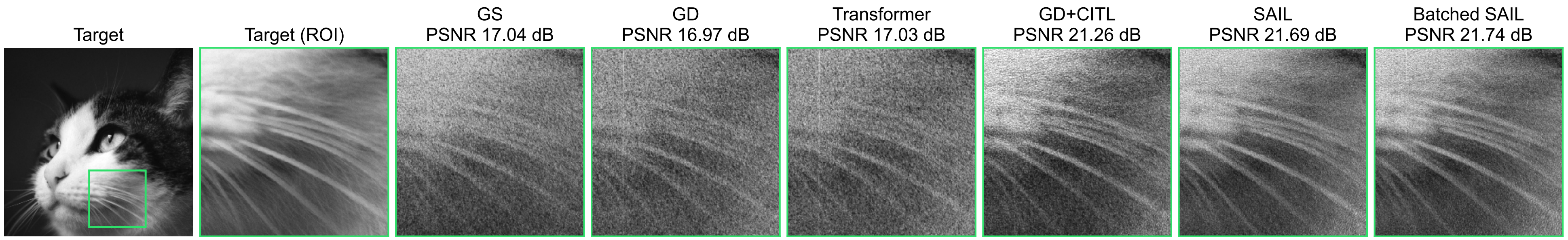

**watering, ideal**

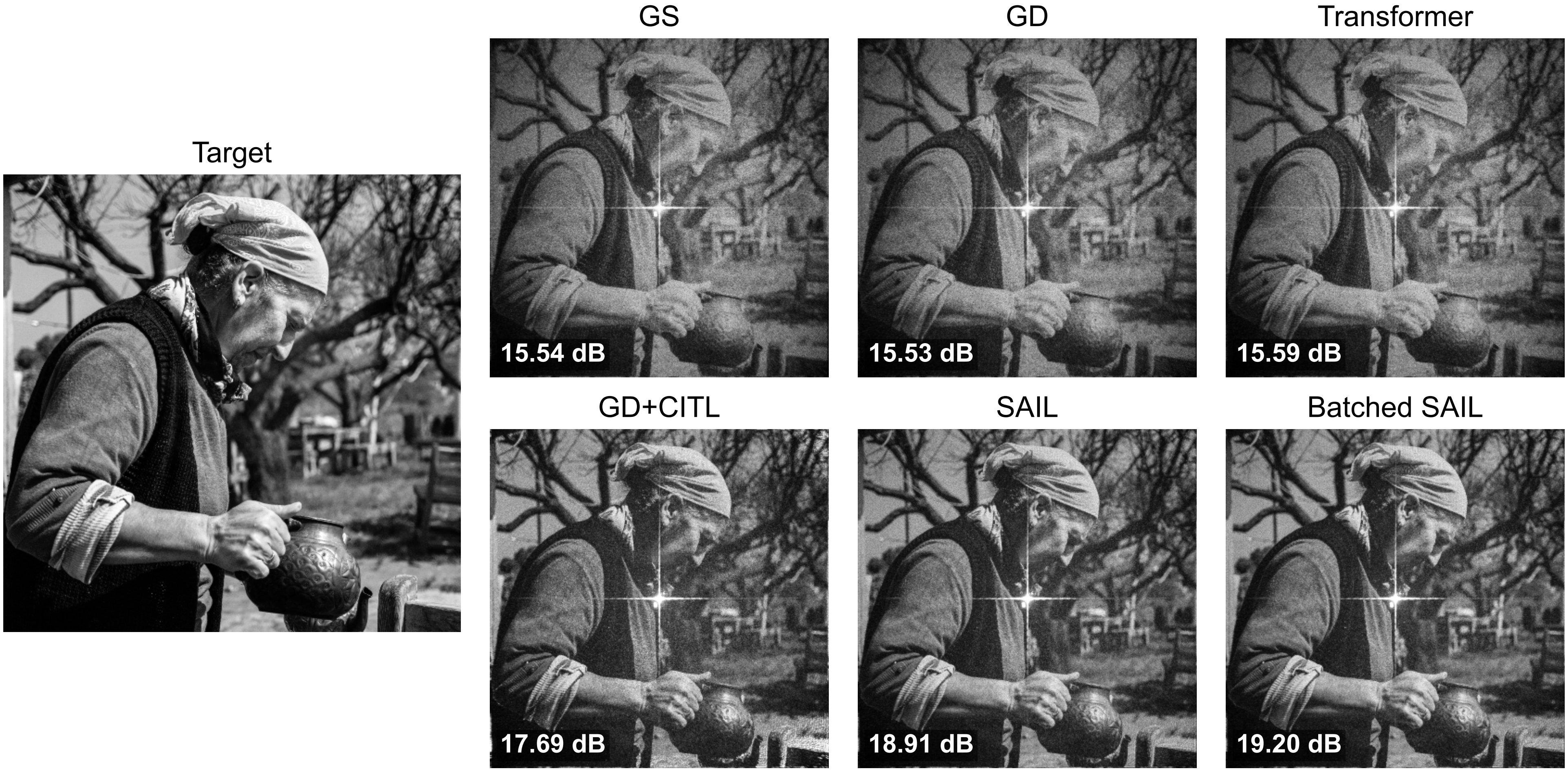

**watering, ideal, regions of interest**

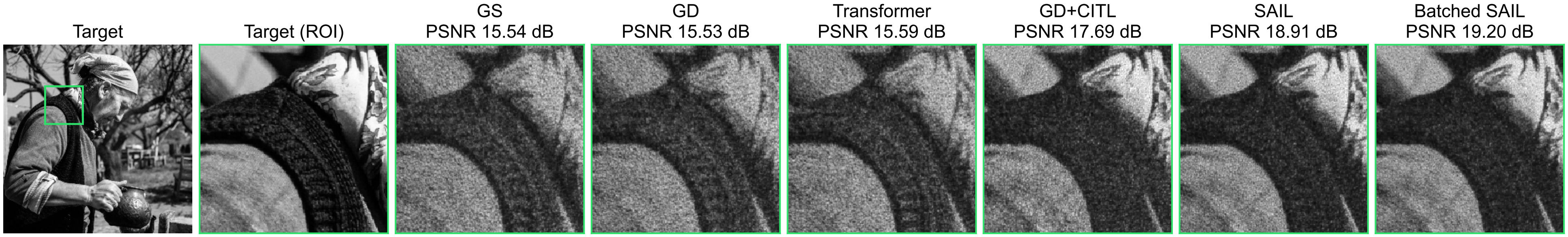

**watering, faithful**

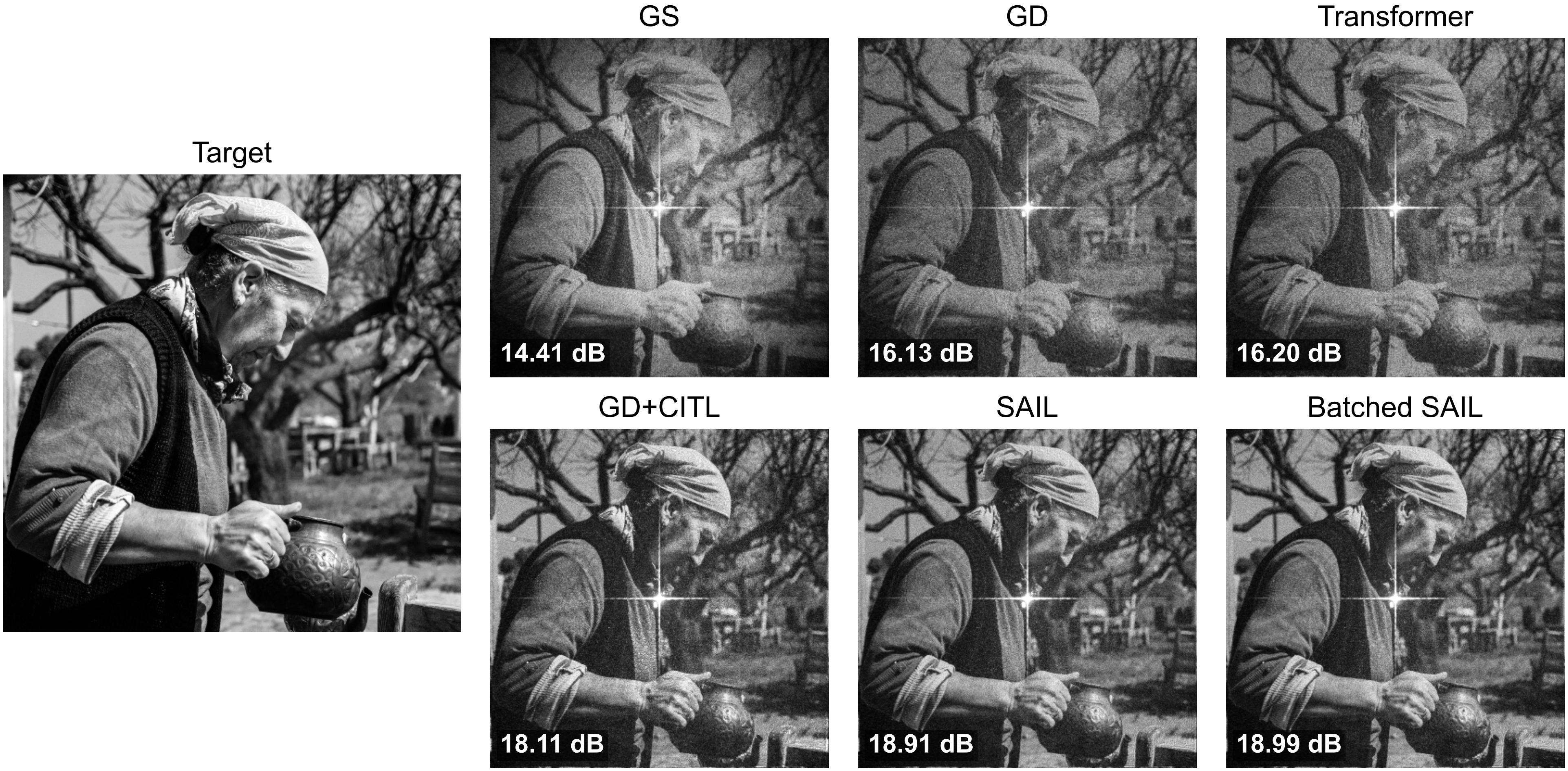

## watering, faithful, regions of interest

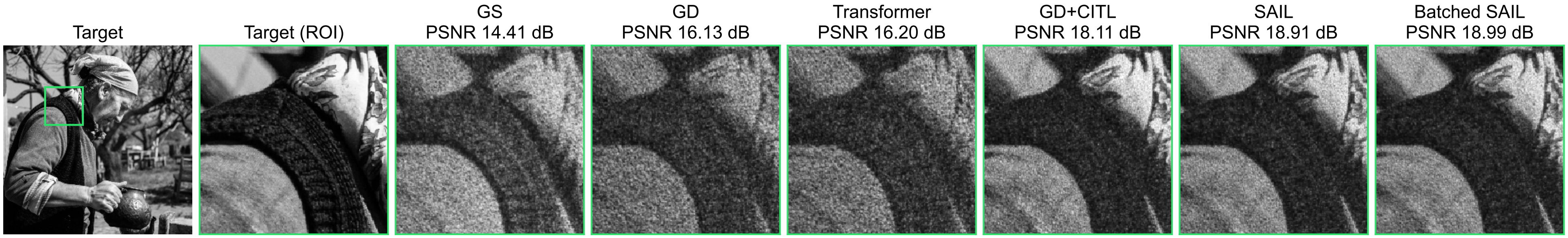

**cat2, ideal**

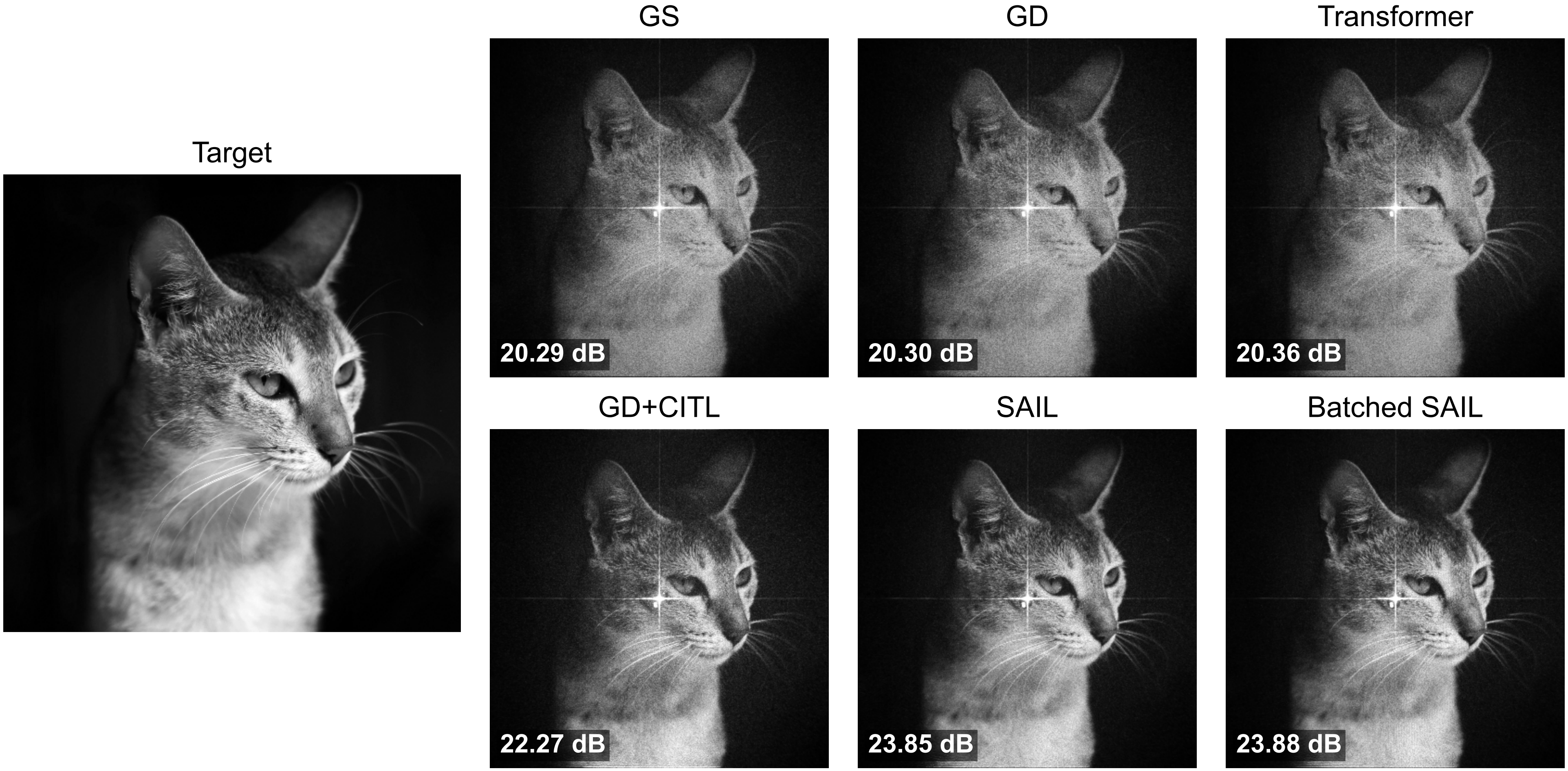
Target
GS
GD
Transformer
GD+CITL
SAIL
Batched SAIL
20.29 dB
20.30 dB
20.36 dB
22.27 dB
23.85 dB
23.88 dB

**cat2, ideal, regions of interest**

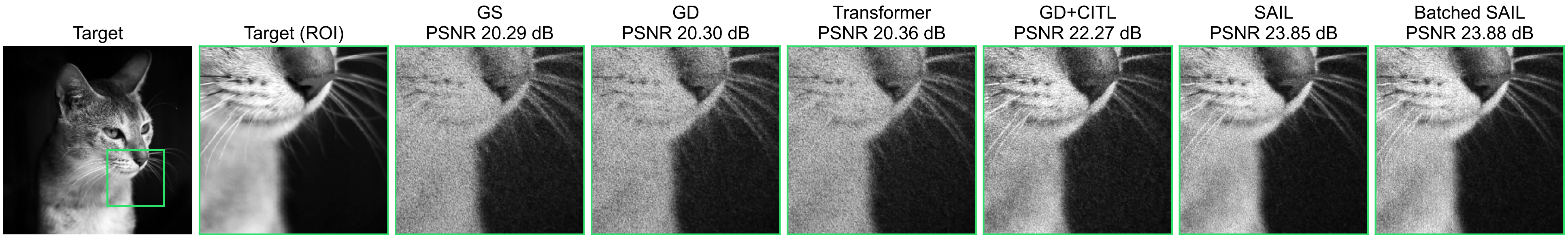

**cat2, faithful**

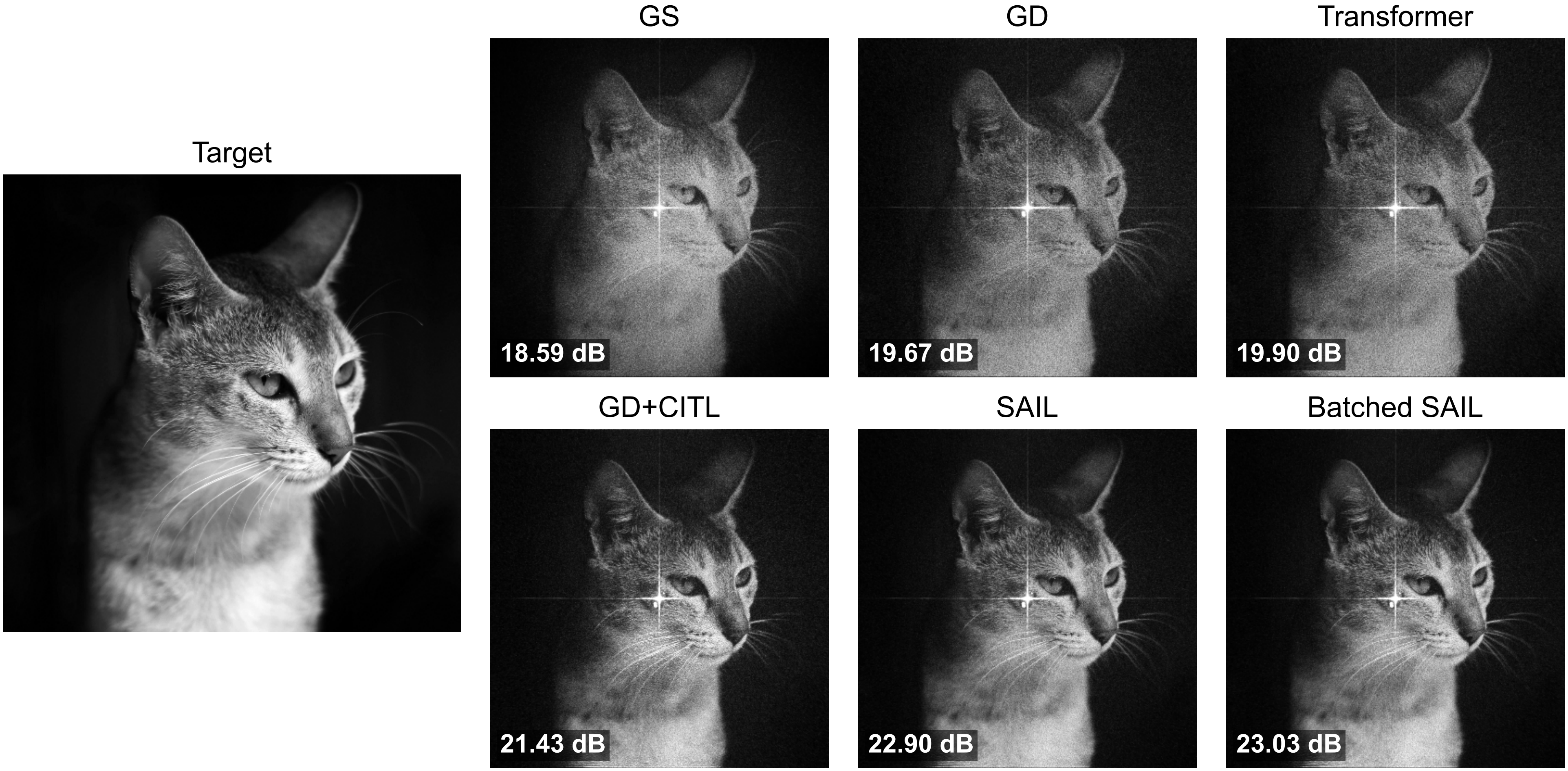

# cat2, faithful, regions of interest

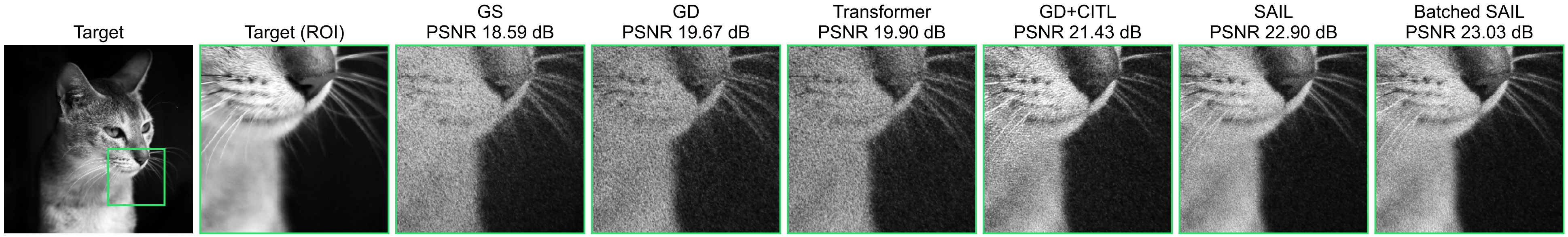

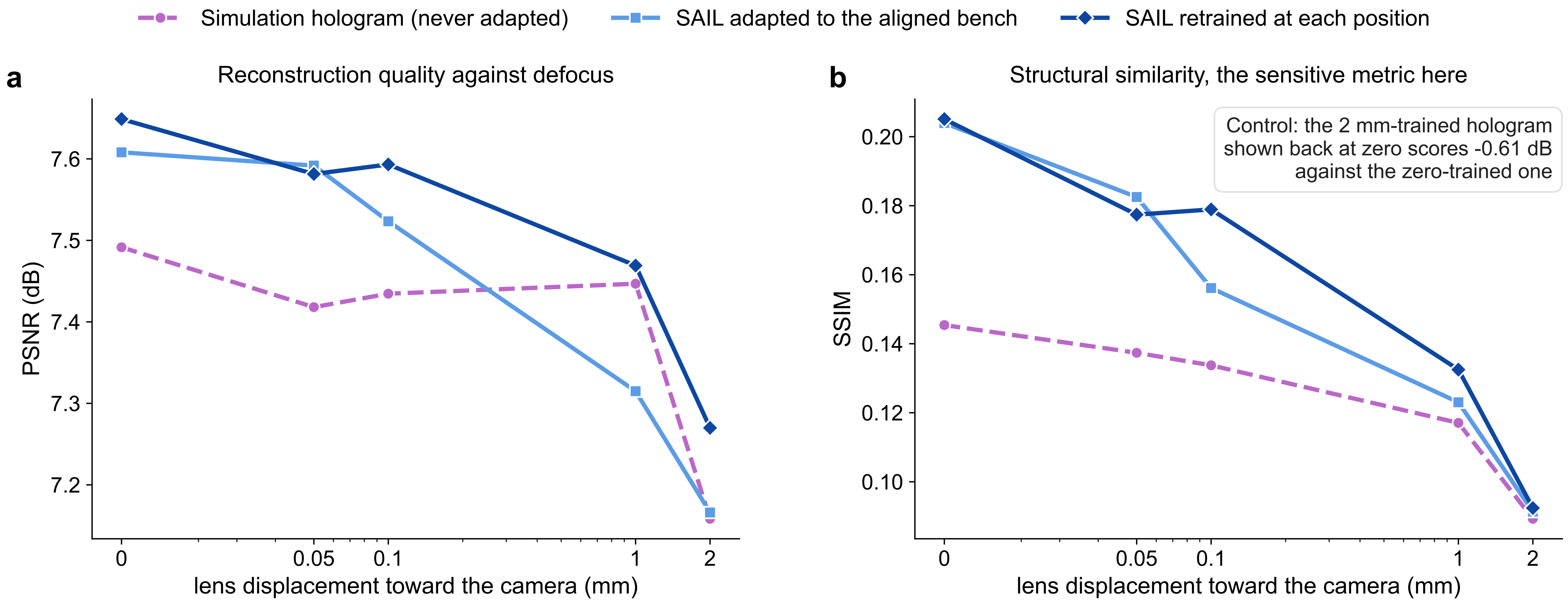
Simulation hologram (never adapted)
SAIL adapted to the aligned bench
SAIL retrained at each position
a
Reconstruction quality against defocus
PSNR (dB)
7.6
7.5
7.4
7.3
7.2
0
0.05
0.1
1
2
lens displacement toward the camera (mm)
b
Structural similarity, the sensitive metric here
SSIM
0.20
0.18
0.16
0.14
0.12
0.10
Control: the 2 mm-trained hologram
shown back at zero scores -0.61 dB
against the zero-trained one
0
0.05
0.1
1
2
lens displacement toward the camera (mm)

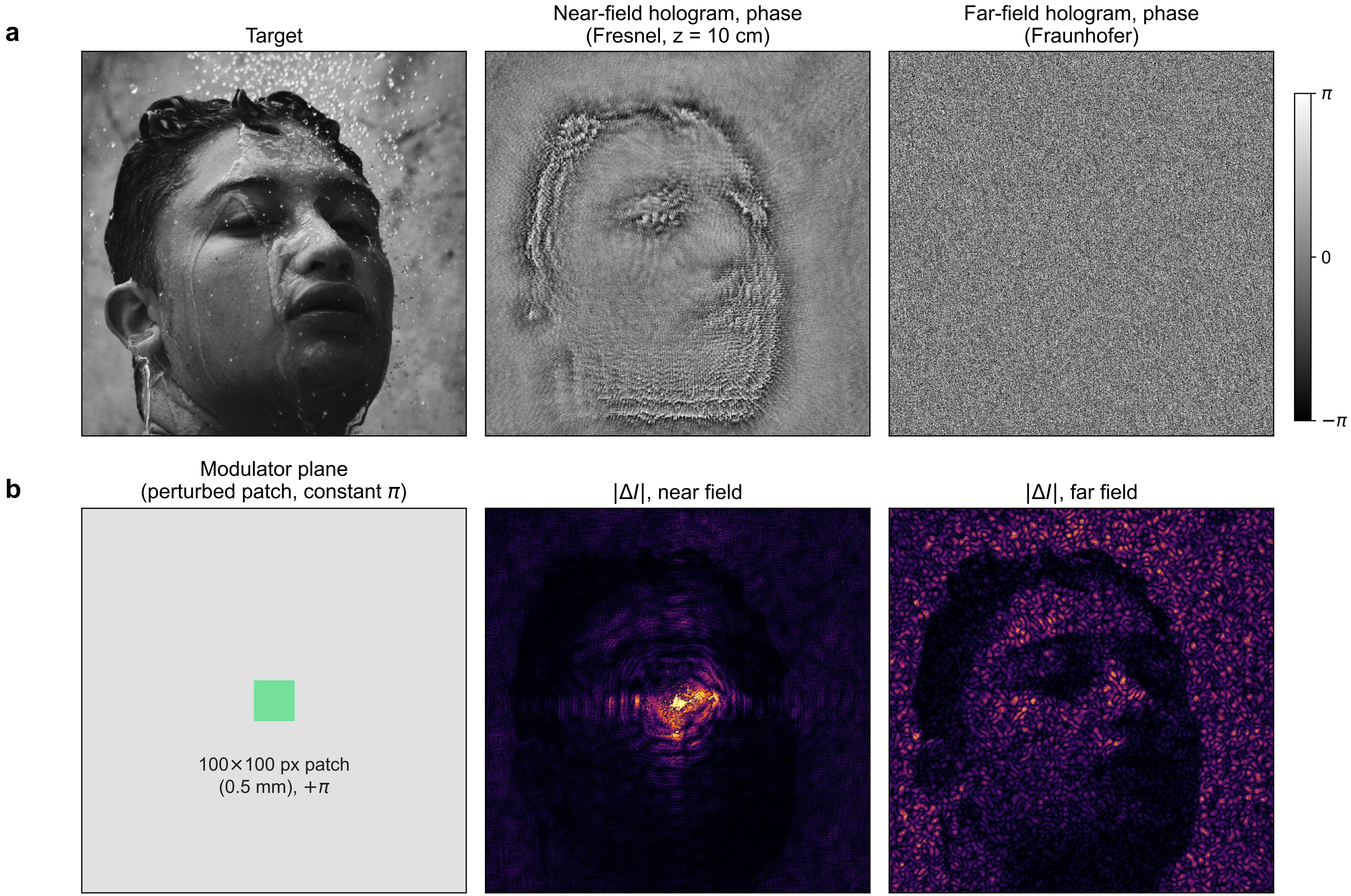

a
Target
Near-field hologram, phase
(Fresnel, z = 10 cm)
Far-field hologram, phase
(Fraunhofer)
π
0
−π
b
Modulator plane
(perturbed patch, constant π)
100×100 px patch
(0.5 mm), +π
|ΔI|, near field
|ΔI|, far field

## Methods

The methodology proceeds from simulation-based training and evaluation of the hologram generator to experimental deployment and hardware-adaptive optimization. We first describe the optical system and forward model, followed by simulation training and benchmarking against classical phase retrieval methods. Simulated results are tested experimentally across methods. We then introduce self-attention-in-the-loop (SAIL) optimization, before describing extensions to misalignment and generalization experiments.

### 1. Optical system and experimental setup

The experimental setup consisted of a reflective liquid crystal on silicon (LCoS) spatial light modulator (SLM) (Holoeye LUNA) illuminated by a single-mode fiber-coupled laser source operating at a wavelength of 650 nm. The laser output was collimated using an achromatic lens (Thorlabs AC254-050-AB-ML, f = 50 mm, broadband anti-reflection coating 400-1100 nm) to produce a uniform incident beam. All results in this work are monochromatic at this single wavelength. Nothing in the training loop is tied to a particular wavelength, but color operation would require wavelength-multiplexed illumination hardware the present system does not include, so we state monochromatic operation as an explicit scope limit.

A linear polarizer was used to align the input polarization to the SLM operating axis. The collimated beam was directed through a 50:50 beam splitter (Thorlabs CCM1-BS013/M), enabling reflection-mode operation of the SLM. The SLM provided phase-only modulation with a resolution of 1920 × 1080 pixels and a pixel pitch of 4.5 μm, operating with 8-bit phase quantization.

The modulated wavefront was propagated through a Fourier lens (Thorlabs AC254-100-A-ML, f = 100 mm, broadband anti-reflection coating 400-700 nm), forming the far-field intensity distribution at the focal plane. The resulting diffraction pattern was captured using a digital camera (Canon EOS Rebel T2i), operated on external power so that no battery changes were required during a capture session, avoiding a perturbation that risks disturbing the alignment. A schematic of the optical setup is shown in Extended Data Fig. E7.

### 2. Camera capture preprocessing

Raw camera captures were processed to ensure consistency with the simulated forward model and target representations. Each grayscale image was first cropped to a fixed region of interest corresponding to the Fourier plane and resampled to the 1000 × 1000 target grid, identically under both forward models. The model-dependent resampling is separate and sits in scoring rather than here. Under the faithful model the simulated field is computed on a grid padded by a factor of two, so the target is upsampled to match it before any metric is formed (Section 3). A background offset was removed by subtracting the minimum intensity value, accounting for the non-zero noise floor of the camera sensor. The resulting image was then normalized by its total energy and scaled by the number of pixels, matching the normalization used in simulation. No temporal or ensemble averaging was performed. All results are reported from single-shot captures.

The undiffracted zero-order (DC) component was handled as a fixed accounting exclusion instead of a mitigation. A circular mask centered on the DC peak, identical for every method, was applied during loss computation. Pixels within this region were excluded from both normalization and loss evaluation, ensuring that optimization focused on the structured diffracted field rather than the dominant central spot. The DC component was retained for visualization and qualitative analysis.

## 3. Forward model, Fourier propagation, parameterization

Wave propagation was modeled under the Fraunhofer approximation, in which the far-field complex replay field is proportional to the Fourier transform of the SLM-plane field. The hologram was treated as a phase-only modulation with unit amplitude, such that the SLM-plane field can be written as,

$$U(x,y) = e^{j\phi(x,y)},$$

where $\phi(x,y)$ denotes the programmed phase[12]. The far-field reconstructed field was computed by the Fourier transform, $\mathcal{F}$, of the SLM-plane field,

$$\tilde{U}(f_x,f_y) = \mathcal{F}\{U(x,y)\},$$

and the reconstructed intensity is obtained from the squared magnitude,

$$I(f_x,f_y) = \left|\tilde{U}(f_x,f_y)\right|^2.$$

In implementation, the transformer predicts a two-channel real-valued representation corresponding to the real and imaginary components of the complex field,

$$U(x,y) = U_{\mathrm{Re}}(x,y) + jU_{\mathrm{Im}}(x,y).$$

This representation is normalized pointwise to unit magnitude to enforce phase-only modulation,

$$\hat{U}(x,y) = \frac{U(x,y)}{|U(x,y)|} = e^{j\phi(x,y)},$$

prior to propagation. An FFT shift is applied to center the zero spatial frequency in the replay plane. For methods producing phase-only holograms, phase maps were converted to complex fields via $U(x,y) = e^{j\phi(x,y)}$, ensuring consistent evaluation under the same forward model. All reconstructed and target intensity distributions were normalized by their total energy and scaled by the number of pixels in the target plane ($H$ = height in pixels and $W$ = width in pixels),

$$I_{\mathrm{norm}}(f_x,f_y) = \frac{I(f_x,f_y)}{\sum_{f_x,f_y} I(f_x,f_y)} \cdot HW$$

ensuring consistent scaling across simulation and experimental measurements.

Two forward models are used throughout, and they differ only in how the modulator plane is sampled and how the finite pixel aperture is treated. The ideal model applies the transform above directly, with no zero-padding, no pixel-aperture envelope and unit fill factor, so the expression for the far-field reconstructed field is unchanged.

$$\tilde{U}(f_x,f_y) = \mathcal{F}\{U(x,y)\}$$

The faithful model is the closer approximation to the physical device. It zero-pads the modulator plane by a factor of two before transforming, applies the sinc envelope of the finite pixel aperture, and uses the measured areal fill factor of 0.91 for the Holoeye LUNA, in three steps.

1. Zero-pad $U(x,y)$ from H × W to $U_{\mathrm{pad}}(x,y)$ = 2H × 2W, with $U(x,y)$ centered in the padded field
2. Transform and shift the padded field so 0 frequency is in the center, $\tilde{U}(f_x,f_y) = \mathrm{fftshift}[\mathcal{F}\{\mathrm{ifftshift}(U(x,y))\}]$, orthonormal

3. Multiply the complex field, $\widetilde{U}(f_x, f_y)$, not the intensity, by the pixel-aperture amplitude envelope, $E(m,n)$, then square it:

$$E(m,n) = \text{sinc}(\sqrt{FF} * v_m)\,\text{sinc}(\sqrt{FF} * u_n)$$

$$v_m = (m - M/2)/M, \qquad u_n = (n - N/2)/N, \qquad M = 2H, N = 2W$$

$$I = |\widetilde{U} \cdot E|^2$$

Two further details of that envelope matter for reproduction. It is applied to the complex field before squaring rather than to the intensity afterwards, which keeps the field and the intensity consistent by construction. And the convention is that a fill factor of one places the first zeros of the envelope at the edges of the padded field, so unit fill corresponds to an active aperture equal to the full pixel pitch.

The symbols in that envelope are as follows. The indices m and n run over the rows and columns of the padded replay grid, which holds $M = 2H$ by $N = 2W$ samples, and $v_m$ and $u_n$ are the corresponding normalized coordinates on that grid, each running from -1/2 to +1/2 so that the centre of the replay plane is the origin. $FF$ is the areal fill factor of the SLM pixel, the fraction of its area that modulates light, measured at 0.91 for the Holoeye LUNA SLM. The square root of $FF$ is the linear fill fraction, the width of the active aperture as a fraction of the pixel pitch, and it is the linear quantity instead of the areal one that the argument of the sinc scales with. Throughout, sinc denotes the normalized form, sinc$= \sin(\pi z)/\pi z$. The envelope is evaluated on the same discrete grid as the transformed field, so $E$ multiplies it sample by sample.

Padding changes the sampling. The modulator pitch is unaltered, so the physical size of the replay field is unaltered too. The padding buys a denser replay grid, which the squared field needs because the intensity carries roughly twice the bandwidth of the field and would otherwise alias. The replay is therefore returned at the padded resolution and is not cropped back, and everything it is compared against is supplied at that same resolution: under the faithful model the target is upsampled by bicubic interpolation to 2H × 2W before any loss or metric is formed, and the camera capture, which arrives on the native grid, is resampled to match in the same way (Section 2).

The two models are otherwise identical, so any difference between them is attributable to the aperture and sampling terms alone. Every condition in this work is reported under both, and the two are never merged. They sample the replay plane differently, so their metrics are separate scoring problems rather than two measurements of one quantity.

Input target images were represented as grayscale intensity distributions with size of $1000 \times 1000$ pixels. Each target was treated as a single-channel image, with normalization applied within the forward model during optimization. To enable efficient modeling of global interactions, the input image was partitioned into non-overlapping patches and embedded into a sequence of tokens. Specifically, the target intensity distribution of size $H \times W$ was divided into non-overlapping patches of size p×p, yielding (H/p)(W/p) tokens used as input to the transformer. The transformer operated on this tokenized representation and produced an output sequence corresponding to the predicted hologram field.

Following prediction, the tokenized output was reconstructed into a full-resolution spatial field through an inverse patching operation. The resulting complex field was then used for forward propagation to compute the reconstructed intensity in the target plane. For experimental deployment, the predicted phase was extracted from the complex field and quantized to an 8-bit representation compatible with the Holoeye LUNA SLM hardware. The phase pattern was spatially arranged to match the native SLM resolution before being displayed.

## 4. Transformer architecture

The hologram generator is a transformer that maps target intensity distributions to phase-only holograms. It operates on a sequence of tokens obtained by partitioning the input image into non-overlapping patches as described above, enabling global interactions across the full spatial extent of the target.

The architecture follows a standard transformer design consisting of stacked self-attention layers with multi-head attention and feedforward sublayers[40,41]. Each flattened patch was linearly projected into a latent embedding space, with a learnable two-dimensional positional embedding added, and processed through a series of attention blocks, allowing global interactions across the full input. The transformer output was then linearly projected back to a patch-wise representation of the complex field.

The transformer was configured with a patch size of $p = 500$, resulting in a small number of tokens for high-resolution inputs. The embedding dimension was set to 256, with 16 attention heads and 4 transformer layers. Each layer included a feedforward network with hidden dimension 1024. Each patch was flattened and linearly projected into a latent embedding space prior to attention, and the transformer output was linearly projected back to a patch-wise representation of the complex field. No dropout was applied, and pre-normalization was used to stabilize training. The use of a large patch size reduces the number of tokens while preserving global spatial structure, which we found to be important for achieving high-fidelity reconstructions in this setting. A schematic of this patch-based representation and reconstruction pipeline is shown in Fig. 1**.**

Scaling behavior follows from this token geometry. At $p = 500$ a $1000 \times 1000$ field is four half-field tokens, so the quadratic cost of attention is negligible. Parameters and compute concentrate instead in the shared linear projections that map each token to and from its $2p^2$-dimensional pixel space, The patch-size sweep of Extended Data Fig. E2 varies p under the ideal forward model and identifies this projection, not attention, as the binding constraint at that width and depth.

Patch size was set by measurement. The size p was varied over 500, 250, 125, and 50 on the same 18 targets at $1000 \times 1000$, with embedding dimension, head count, layer count, feedforward width, optimizer, learning rate and epoch budget all held at the values above, and one network trained per target per setting under the ideal forward model, 72 runs in total. Because the linear projections map each token to and from its $2p^2$ pixel values, parameter count is not constant along the sweep. It falls from 195,660,320 at $p = 500$ to 51,288,392, 15,206,930, and 5,186,696 at $p = 250$, 125, and 50. Per-setting medians and the per-target monotonicity check are in Extended Data Fig. E2.

Attention selectivity is measured on trained checkpoints as $1 - H/\log T$, where $H = -\Sigma_t a_t \log a_t$ is the Shannon entropy, in nats, of the attention weights $a_t$ that one head assigns to the T tokens, and $\log T$ is the entropy of the uniform distribution over those same tokens and therefore the maximum H can take. Normalizing by $\log T$ makes the measure independent of token count, so 0 is uniform mixing and 1 is a head attending to a single token, whatever T is. It is computed for every head in every layer on each of the 18 targets, in four regimes: per-target and batched, each trained in simulation and through the physical optics. A fifth condition, the batched network trained through the optics for 750 rather than 2,000 epochs, is computed as a training-length control and is reported in the deposited diagnostics record. Architecture, token geometry, parameter count and data format are identical across the four, so the training regime is the only variable. Selectivity averaged over heads, layers and targets is reported in Fig. 5c. The attention maps it is computed from are shown one per regime in Extended Data Fig. E3a and target by target, on the same scale, in Extended Data Fig. E3c, so that the average can be checked against the distribution it summarizes. The bias-dominance control evaluates the same checkpoints with the input-dependent pathway removed, so that only the learned bias reaches the output, and scores the result with the metric pipeline used everywhere else (Extended Data Fig. E3b). Both diagnostics evaluate checkpoints through the simulated forward model and are therefore reported in simulation only.

**5. Simulation training and baseline**

Using this forward model, we first trained and evaluated the hologram generator in simulation under the idealized Fraunhofer forward model described in Section 3. For the main simulation experiments, a separate transformer was trained for each of the 18 target images at a resolution of $1000 \times 1000$ pixels. Training was performed with batch size 1 for 10,000 epochs using the AdamW optimizer[50] with betas of 0.9 and 0.999, a weight decay of 1e-4, and an initial learning rate of $10^{-4}$. The objective was the mean squared error between the reconstructed intensity predicted by the differentiable Fourier model and the normalized target intensity. The learning rate was decayed multiplicatively by a factor of 0.95 every 1,000 epochs. The best-performing checkpoint, as determined by the training loss, was retained for evaluation. The best checkpoints concentrated in the final 1% of the training budget (minimum/median/maximum best epoch 9935/9992/9999 across the 18 targets, all above epoch 9900, with validation loss still falling), so the matched-budget comparison treats the transformer and the iterative baselines symmetrically: neither had stopped improving when its budget ended. Convergence behavior for all methods is reported in Extended Data Fig. E5. Implementation settings for every condition are tabulated in Table 3, generated programmatically from each run's archived configuration snapshot with a consistency check across runs. Campaign-level compute is reported in Table 2.

To benchmark the learned approach against classical hologram synthesis methods, we compared it with GS and GD[18-20]. GS was implemented as alternating projection, with the target enforced as a replay-plane amplitude constraint at each iteration. GD as direct optimization of the modulator phase against a loss on the recovered intensity, through the same differentiable forward model used throughout the paper. Both implementations are inspectable in the code release. For both baselines, the target image was parameterized as a target amplitude, the square root of the desired intensity. This parameterization does not set the loss domain. For GD the implementation squares the amplitude back to intensity before the loss is formed, so the objective is a mean-squared error in the intensity domain. For GS the amplitude enters as the replay-plane magnitude constraint of alternating projection, the classical amplitude-domain standard[18,20]. The resulting hologram phase was evaluated under the same Fourier replay model used for the learned approach. GS and GD were initialized with random phase and evaluated across iteration sweeps to characterize convergence behavior. In the reported simulation comparison, both methods were swept over 10, 25, 50, 75, 100, 200, 250, 300, 500, 750, 1,500, 3,000, 5,000, 7,500 and 10,000 iterations, and every budget the sweep passed through was written to disk, which is what makes the 750-iteration operating point and the 10,000-iteration converged arm the same run rather than two.

To confirm that the amplitude parameterization is not a favorable choice made on the baselines' behalf, GS and GD were re-run with the target supplied as raw intensity in place of amplitude, on all 18 targets under both forward models, with every other setting unchanged and with scoring by the same pipeline. Differences are paired by target and tested with a two-sided

Wilcoxon signed-rank test over the 18 pairs. Significance is read from the saved summary table. The comparison is reported in Extended Data Fig. E4, and both arms appear in Table 1.

For GD, optimization was performed with the Adam optimizer[51], a learning rate of 0.03, and an intensity-based loss, with no total variation regularization[52], low-pass filtering, or carrier term. GS was evaluated under the corresponding multilevel phase setting, likewise without low-pass filtering or carrier modulation. Iteration 750 was used as the evaluation point for GS and GD. These choices were made to compare all methods under a common Fourier-optical setting without introducing additional hand-tuned regularization terms. For the transformer baseline in the simulation benchmark, a pretrained per-target checkpoint was loaded for each image and evaluated in a single forward pass. Reconstruction quality was quantified using PSNR and NMSE (with additional metrics including SSIM, MSE, and diffraction efficiency computed but not reported in the main text), all evaluated after replay-field normalization. The zero-order mask of Section 2 was applied identically for every method during scoring. Masked pixels contribute neither to the normalization nor to any reported metric, and the component is retained for visualization only. Wall-clock time was also recorded to compare the learned single-pass inference with iterative optimization. The 750-iteration operating point is three times a typical published far-field phase-retrieval budget, and extended sweeps to 10,000 iterations are reported in Extended Data Fig. E5. Full implementation settings for every method, including the hardware and software environment and compute, are tabulated in Table 3.

In addition to per-target training, we evaluated a batched transformer trained jointly across all target images. In this setting, a single network was optimized over multiple targets simultaneously using a batch size of 18, one batch covering every target, with each batch containing distinct images. This contrasts with the per-target configuration, where a separate transformer is trained for each image with batch size 1. The batched transformer was then evaluated without further adaptation on the individual targets, demonstrating that a single network can predict phase holograms for all targets at comparable performance to the per-target transformers.

As an architectural control we trained an FNO per target under the ideal forward model. The operator is the standard published formulation, unmodified. It used 500 × 500 Fourier modes, which is full coverage of the 1000 × 1000 grid rather than the usual mode truncation, width 8, four operator layers, projection width 128 and GELU activations, giving 256.0 million parameters (Table 3d). Two arms were run. The from-scratch arm optimized the same intensity-domain objective through the same differentiable forward model used for the transformer, for 10,000 epochs at a learning rate of 0.01 with weight decay 1e-4 and a multiplicative decay of 0.95 every 1,000 epochs. Because a failure to train can be an artifact of the learning rate, that rate was separately probed over five orders of magnitude, from 1 to 1e-4, on two targets at a reduced 1,000-epoch budget. That probe is a hyperparameter study instead of part of the canonical dataset, so it is summarized by the mean over those two targets, which moved only between 8.61 and 8.99 dB across the five rates. The outcome is not a tuning failure. The regression arm removed the inverse problem entirely. The far-field objective was replaced by a mean-squared error between the operator's output and GD's own converged phase, both normalized to unit modulus and compared as (cos, sin) pairs so that only the phase is scored and the 2π wrap introduces no discontinuity, for 1,500 epochs, within which it converged. The network, optimizer and budget are otherwise those of the from-scratch arm, so the two differ only in what they are asked to fit. The operator variant was selected in simulation, never under camera feedback, since sweeping on the bench is not affordable. The operator is run under the ideal forward model only because its quality leaves no room for the faithful model to change the conclusion. Both arms are scored by the same pipeline as every other method. Results are in Extended Data Fig. E1 and Table 1, settings in Table 3d, and the runs are in the deposited record.

### 6. Benchmark comparison experiments

To evaluate how simulation-trained methods translate to the physical optical system, we performed an experimental comparison across a set of 18 target images. For each target, holograms were generated using GS, GD, the pretrained per-target transformer, and a batched transformer trained jointly across multiple targets, and were evaluated sequentially under identical conditions. Iteration 750 was used for GS and GD as the comparison point. The 10,000-iteration phases from the same sweep were displayed and photographed in the same session and are reported in Table 1. Figure 3 uses the 750-iteration operating point for both baselines, so that every comparator on that figure sits at the budget the main text reports, and the converged arms are read from Table 1 and Extended Data Fig. E5 instead.

For each method, the corresponding phase hologram was displayed on the SLM, and the resulting far-field intensity distribution was recorded using a single camera capture. Exposure settings were determined using the GS reconstruction for each target and then held fixed across all methods to ensure consistent intensity scaling.

For the bench comparison, every hologram from every method was re-displayed and re-photographed in a single session on a single alignment, and every reported bench value is scored from that session. The original session captures are retained in the deposited record alongside it, so the session-to-session difference is recoverable. Re-photographing the same stored holograms in a separate session on a separate alignment shifts scores by a median of 0.18 dB, with a per-target range of -1.01 to +0.05 dB, while photographing identical holograms on two dates eleven days apart, as a rig-reproducibility check, gives a median absolute difference of 0.14 dB. The negative drift is carried by the camera-adapted methods, in proportion to how much each is adapted to the optics, so the bench values reported here are the conservative ones.

All captured images were processed using the pipeline described in Section 2. This experimental comparison established the gap between simulation-based predictions and physical reconstructions, motivating the use of self-attention-in-the-loop (SAIL).

### 7. SAIL optimization

To bridge the gap between simulation and experiment, we incorporated SAIL, enabling the network to adapt directly to the physical optical system. In this framework, holograms generated by the transformer were displayed on the SLM, and the resulting intensity distribution was captured by a camera and used as feedback during training. SAIL trains the generator’s weights against the measured optics. It is not a refinement stage applied to a fixed hologram, and no hologram produced by any other method enters the loop.

For each training iteration, the network predicted a phase-only hologram, which was quantized to an 8-bit representation and displayed on the SLM. After a brief settling period, the corresponding optical reconstruction was captured using the camera. The captured image was preprocessed as described in Section 2.

To enable gradient-based optimization through the non-differentiable hardware loop, we employed a straight-through estimator in which the forward signal is replaced with the experimentally measured intensity while gradients are computed using the simulated replay field. This was implemented as

$$I_{\text{proxy}} = I_{\text{sim}} + \text{sg}(I_{\text{cam}} - I_{\text{sim}}),$$

where sg denotes a stop-gradient operation, which prevents gradients from propagating through its argument during backpropagation. In the forward pass, this reduces to $I_{\text{proxy}} = I_{\text{cam}}$, ensuring that the loss is evaluated using the experimentally measured intensity, while gradients in the backward pass flow through the simulated intensity, $I_{\text{sim}}$, enabling optimization through the differentiable model.

The optimization objective minimizes the discrepancy between the reconstructed and target intensities,

$$\mathcal{L} = \frac{1}{HW} \sum_{x,y} \left( I_{\text{proxy}}(x, y) - I_{\text{target}}(x, y) \right)^2$$

with an additional simulated loss term included during training to stabilize optimization. The full objective is L = L_cam + λ_sim L_sim with λ_sim = 0.1, held constant for every SAIL condition reported here (Table 3a). The camera term is what adapts the network to the physical system. The simulated term is a consistency constraint: it holds the solution near one the differentiable forward model can also explain, so that the network is not free to fit whatever a single capture reports, including its noise and its exposure. The weight sits an order of magnitude below the camera term so that measurement leads and simulation regularizes. The experimental loss incorporated the camera measurement via the straight-through proxy. This formulation allowed the network to retain the stability of simulation-based gradients while adapting to system-specific deviations captured by the hardware.

Per-target SAIL was performed with batch size 1, reflecting the sequential nature of hardware acquisition. Batched SAIL, instead, passes over all 18 targets within each epoch, one exposure each (Algorithm 2). Each per-target network was trained for 750 epochs with the same optimizer as the simulation training, AdamW with betas of 0.9 and 0.999 and a weight decay of 1e-4, and with exposure settings fixed for each target image and held constant throughout training to ensure fair comparison. The learning rate was decayed by the same factor of 0.95 but every 250 epochs instead of every 1,000 because the CITL budget is 750 epochs and a 1,000-epoch period would leave the rate at its initial value for the entire run. Both schedules are tabulated in Table 3a. This procedure enabled stable optimization while maintaining consistency between simulation and experimental evaluation.

**Algorithm 1 | SAIL, per-target training.** Step comments key each line to the numbered walkthrough of Fig. 1. HALO is the network of Section 4 and returns the complex field directly. The Fraunhofer operator on line 7 is the forward model of Section 3, ideal or faithful. I_proxy takes its value from the camera and its gradient from the simulated model. Lambda_sim = 0.1, constant. The reported hologram is the checkpoint with the lowest camera loss.

Input: target intensity I_tgt (H x W), patch size p, epochs E, weight lambda_sim, and learning rate eta
Output: phase hologram phi*

```
 1  theta <- initialize HALO  (hologram generator class)                    # fresh per target
 2  best <- infinity
 3  for e = 0 ... E-1:
 4    if e > 0 and e mod 250 == 0: eta <- 0.95*eta
 5    X    <- patchify(I_tgt, p)                    # Fig. 1, steps 1-2
 6    Y    <- HALO(X)                               # the network; Fig. 1, steps 3-7
```

```
 7    I_sim <- normalize(|Fraunhofer{Y}|^2)          # forward model, Section 3;
 8    phi   <- arg(Y)                                # Fig. 1, step 8
 9    display(quantize_8bit(phi)) on the SLM; settle # Fig. 1, steps 9a-10
10     frame <- camera.capture()                     # Fig. 1, step 11
11     I_cam <- calibrate(frame)                     # ROI, de-rotate, resample,
                                                     # subtract floor, normalize
12     M     <- 1 - DC_mask                          # zero order excluded
13     I_proxy <- I_sim + stopgrad(I_cam - I_sim)    # Fig. 1, step 12
14     L_cam <- mean_M((I_proxy - I_tgt)^2)
15     L_sim <- mean_M((I_sim - I_tgt)^2)
16     L     <- L_cam + lambda_sim*L_sim
17     theta <- AdamW(theta, grad_theta L, eta)      # Fig. 1, step 13
18     if L_cam < best: best <- L_cam: phi* <- phi
19  return phi*
```

**Algorithm 2 | Batched SAIL.** One shared network serves every target. Batched training repeats steps 9a-13 of Fig. 1 once per target per epoch with a single shared theta.

```
 1  theta <- initialize HALO (hologram generator class)         # ONE network for all targets
 2  for e = 0 ... E-1:
 3    for t in targets:              # one exposure per target per epoch
 4      lines 5-17 of Algorithm 1 with I_tgt = t
 5    if mean camera loss improved: checkpoint theta
 6  return theta                     # any target: one forward pass
```

**8. Misalignment experiment**

To evaluate robustness to deviations in the optical system, we introduced a controlled misalignment by shifting the Fourier lens along the optical axis toward the camera by nominal caliper displacements between 0.05 and 2 mm. Such perturbations alter the effective propagation distance (defocus) and introduce phase distortions that degrade reconstruction quality when unaccounted for[53]. Displacements are nominal caliper settings with an uncertainty of approximately ±0.05 mm.

Following this perturbation, the system was evaluated under the same experimental conditions as described above. The transformer without experimental adaptation exhibited degraded performance under misalignment, reflecting its sensitivity to changes in the forward model.

To assess the ability of the proposed framework to compensate for such deviations, SAIL was implemented from random initialization under the misaligned configuration. All other experimental parameters, including exposure settings and preprocessing, were held fixed. The lens was returned to its nominal position and re-verified as a bookend at the end of the sweep. As a converse control, the hologram trained at the 2 mm displacement was displayed on the realigned system, so that one hologram is scored under an optical state it was not trained on (Extended Data Fig. E9). Those return-to-zero reconstructions scored within +0.03 dB (SAIL) and +0.04 dB (simulation-only) of the aligned baseline, both well inside the rig's day-scale reproducibility of 0.14 dB.

This procedure allowed the network to adapt directly to the altered optical system. The scope of that adaptation is characterized by the aberration-severity study (Extended Data Fig. E9). Adaptation is specific to the optical state under which it was trained, and it does not restore quality that the aberration has physically removed.

## 9. Generalization experiments

To evaluate whether the learned mapping extends beyond the training distribution, we assessed generalization on low-resolution target patterns using a separate set of experiments based on $28 \times 28$ intensity images. In this setting, the network was trained under a constrained curriculum that included only a subset of target classes, allowing explicit evaluation of out-of-distribution (OOD) performance.

In particular, training was restricted to three digit classes, 2, 6 and 9, from the MNIST dataset[43], alongside a mixture of synthetic patterns including structured primitives, sparse and dense random patterns, and band-limited textures. Despite this limited supervision, the trained network was evaluated on all ten MNIST digit classes, as well as on manually constructed target patterns not present in the training data. The network used for this study is a compact 3.3-million-parameter proof-of-concept model, separate from the main-text network and matched to the 28 × 28 grid, with a patch size of 14 so that the field is again four tokens and the attention matrix again 4 × 4, as in the main-text architecture.

The network produced recognizable reconstructions for unseen digits and custom patterns, preserving global structure despite the strict phase-only constraint and limited spatial resolution. Generalization performance was further assessed on hand-designed targets with irregular spatial structure, demonstrating the ability of the network to synthesize coherent far-field intensity patterns beyond the distribution of training examples.

These results indicate that the self-attention-based hologram generator learns a transferable inverse mapping between target intensity and phase modulation, rather than memorizing individual training instances. This behavior is consistent with the global coupling imposed by Fourier propagation, where coherent structure must be maintained across the entire field. Representative reconstructions, each shown as the desired target far field, the reconstructed far field and the predicted phase, are in Extended Data Fig. E6, and the archived outputs are in the deposited record.

## 10. Regime comparison simulation

The comparison in Extended Data Fig. E10 is an illustrative simulation, and no quantity from it enters any reported result. Both arms used the same Gerchberg-Saxton alternating projection, the same 60 iterations and the same zero-phase initialization on a 1000 × 1000 modulator at the rig's 4.5 μm pitch under 650 nm illumination, so that propagation was the only difference between them. The grid, pitch and wavelength are stated because the near-field arm needs them: the angular spectrum transfer function is a function of wavelength, propagation distance and sample spacing, so the comparison is not reproducible without them. The near-field arm propagated with the angular spectrum method at z = 10 cm, applying the angular spectrum transfer function with evanescent components excluded on the forward path and its conjugate on the reverse path. That transfer function is the exact scalar propagator instead of the Fresnel paraxial approximation, so the spatial locality the near-field arm exhibits is not an artifact of a paraxial expansion. The far-field arm used the single Fourier transform of the modulator field on both paths.

Locality was probed by adding a constant π offset to a centered 100 × 100 pixel patch of the converged hologram and imaging the resulting change in reconstructed intensity. A randomized patch was deliberately not used. A random patch would scatter light across the modulator's entire angular bandwidth whichever regime it sat in, so the resulting spread would be set by the pixel pitch and would look the same in both. A constant offset keeps the disturbance narrow-band, which is what allows the propagation, rather than the sampling, to decide where the change lands. Concentration is reported as the fraction of that change falling within a box three patch widths across and centered on the patch, against the 9% expected if the change were spread uniformly over the field.

### 11. Implementation and compute

All networks, the classical baselines and the differentiable forward models are implemented in PyTorch and run in single-precision floating point on one NVIDIA RTX 4000 Ada Generation laptop GPU with 12 GB of memory, in a workstation with an Intel Core i9-13950HX processor and 64 GB of system memory, under Python 3.14.4 with NVIDIA driver 596.58. Gerchberg-Saxton, gradient descent, the FNO and the self-attention generator all execute on that same device through a common evaluation harness, so every wall-clock figure reported in this work is directly comparable. The 18 per-target SAIL runs total 6.82 hours, the batched sessions take 8.35 hours at 750 epochs and 23.03 hours at 2,000 epochs, and inference is a single forward pass of 12 to 24 ms. The random-initialization arms of the iterative baselines are seeded, and the zero-phase arms are deterministic by construction. Wall-clock figures were measured on a workstation that was in intermittent use for other tasks during the campaign, which is why inference is quoted as a range. Every network reported here, including the 195.7-million-parameter batched one, trains within the 12 GB of this single device, so no result in this work requires cluster-scale hardware.

### 12. Statistics and reporting

Every quantity in this work is computed over the same 18 targets, and no target is excluded from any comparison. The narrative quotes the median across those 18 because the distributions are skewed at this sample size. Means and standard deviations travel beside medians and interquartile ranges in Table 1, so both are available. Comparisons between methods are paired by target. The reported difference is the median of the 18 per-target differences, its uncertainty is a 95% confidence interval from a bootstrap with a fixed seed, and significance is a two-sided Wilcoxon signed-rank test over the same 18 pairs. Wins are the number of targets on which the difference favors the first method. No correction for multiple comparisons is applied, and P values are reported per comparison. The ideal and faithful forward models are separate scoring problems, at different sampling, so they are reported side by side and never merged or compared against one another. Every metric is computed after replay-field normalization with the zero-order mask of Section 2 applied identically to every method, by a single scoring pass over the whole dataset, so that no number in this work is produced by a method-specific script.

### Method references

Numbering continues from the main reference list. References cited in both the main text and the Methods retain their main-list numbers.